# Stress-Induced Ferroelectricity in Hafnia Core-Shell Nanoparticles

Anna N. Morozovska[1*], Eugene A. Eliseev[2], Richard (Yu) Liu[3],
Sergei V. Kalinin[3†], and Dean R. Evans[4‡]

[1] Institute of Physics of the National Academy of Sciences of Ukraine, 46, Nauky Avenue, 03028 Kyiv, Ukraine

[2] Frantsevich Institute for Problems in Materials Science of the National Academy of Sciences of Ukraine, 3, str. Omeliana Pritsaka, 03142 Kyiv, Ukraine

[3] Department of Materials Science and Engineering, University of Tennessee, Knoxville, TN, 37996, USA

[4]Zone 5 Technologies, Special Projects, San Luis Obispo CA 93401, USA

**Abstract**

In contrast to hafnia ($HfO_2$) thin films, where the appearance of switchable ferroelectric polarization can be induced by strain or defect engineering, reliable methods for controlling ferroelectricity in $HfO_2$ nanoparticles are absent. Direct experimental observations of ferroelectric hysteresis and ferroelectric domains in these nanoparticles have also not been reported. To the best of our knowledge, stress-induced ferroelectric states in the $HfO_2$ nanoparticles have not been explored. In this work, we study the influence of chemical stress on phase diagrams, dielectric properties, and polar properties of spherical $HfO_2$ core-shell nanoparticles using a Landau-Ginzburg-Devonshire free energy functional that includes trilinear and biquadratic couplings involving polar, antipolar, and nonpolar order parameters. The ferroelectric phase exhibits reentrant behavior as a function of nanoparticle size, such that the spontaneous polarization exists only within a limited range of core radii $R_c$, namely $R_{cr}^{min} < R_c < R_{cr}^{max}$. The minimal critical radius $R_{cr}^{min}$ is primarily determined by the size dependence of the depolarization field and correlation effects, whereas the maximal critical radius $R_{cr}^{max}$ is governed mainly by the size dependence of chemical stresses induced by the elastic defects in the shell. Thus, this work identifies a stress-driven mechanism for reentrant ferroelectricity stabilization in nanoscale hafnia systems, arising from the competition between depolarization field-induced suppression of ferroelectricity and its stabilization by shell-induced chemical stress. We revealed that relatively large compressive chemical strains are necessary to induce the ferroelectric phase in the $HfO_2$ nanoparticles. Successful chemical strain engineering opens the way for significant enhancement of nanoscale $HfO_2$ polar properties for applications in advanced memory cells and logic devices.

[*]Corresponding author, e-mail anna.n.morozovska@gmail.com

[†] Corresponding author, e-mail sergei2@utk.edu

[‡] Corresponding author, e-mail: deanevans@zone5tech.com

## I. INTRODUCTION

Thin films of hafnia-zirconia ($Hf_xZr_{1-x}O_2$) are among the most promising silicon-compatible ferroelectric materials for advanced electronics memories [1, 2, 3, 4]. At the same time, they are one of the more enigmatic ferroelectric systems, since the physical mechanisms responsible for the emergence of ferroelectric and/or antiferroelectric behavior remain under active debate [5, 6]. The transition from the bulk nonpolar monoclinic m-phase (space group *P2₁/c*) to the ferroelectric (FE) orthorhombic o-phase (space group *Pca2₁*) can occur in $Hf_xZr_{1-x}O_2$ films when the thickness $h$ decreases below a critical value $h_{cr} \cong 20-30$ nm [7], whereas in more typical ferroelectric materials the FE phase becomes stable at larger thicknesses, $h > h_{cr}$. According to direct experimental observations [8] and ab initio calculations [9, 10], the switching path of spontaneous polarization $\vec{P}_s$ in $Hf_xZr_{1-x}O_2$ thin films is indirect, in that the transition from $+\vec{P}_s$ to $-\vec{P}_s$ states proceeds via the nonpolar tetragonal t-phase.

The existence of ferroelectricity and the mechanism of polarization switching are much more complicated in small $Hf_xZr_{1-x}O_2$ nanoparticles with $x \geq 0.5$ and sizes 10 – 30 nm. Due to their small sizes, it is difficult to distinguish between the t-phase (space group $P4_2/nmc$) and the o-phases (space groups *Pbca*, *Pbcm*, and ferroelectric *Pca21*) using X-ray diffraction (XRD) analysis, since the corresponding peaks are very close and broadened [11]. In contrast to $Hf_xZr_{1-x}O_2$ thin films, where switchable ferroelectric polarization can be induced by strain or defect engineering [12, 13, 14, 15, 16], reliable methods of controlling ferroelectricity are absent in $Hf_xZr_{1-x}O_2$ nanoparticles. Furthermore, direct experimental evidence of ferroelectric hysteresis and/or ferroelectric domains in the nanoparticles has not been reported. Since the coupling between polar and nonpolar modes enhanced by large compressive strains can induce robust ferroelectricity in $HfO_2$ thin films and nanoislands [17], it is therefore plausible that similar effects may also occur in $Hf_xZr_{1-x}O_2$ nanoparticles. However, a theoretical framework describing stress-induced ferroelectric states in nanoparticles has not yet been developed.

Several experimental [18, 19] and theoretical [20, 21] studies have revealed a leading role of oxygen vacancies [22, 23] in the emergence and stabilization of the FE o-phase in $Hf_xZr_{1-x}O_2$ thin films. Thermodynamic considerations [24] based on the Landau-Ginzburg-Devonshire (LGD) approach predict that ferro-ionic states can be stable in small $Hf_xZr_{1-x}O_2$ nanoparticles due to high degree of screening provided by ionic-electronic charges. The prediction was later supported experimentally by the observation that small (size 8 – 10 nm) oxygen-deficient $Hf_xZr_{1-x}O_2$ nanoparticles, which contain a large fraction of indistinguishable o-phases due to the annealing in a reducing $CO+CO_2$ atmosphere [25, 26], can exhibit ferroelectric-like properties, such as a colossal dielectric response over a wide frequency range [27], as well as demonstrate resistive switching and pronounced temperature-dependent charge accumulation [28, 29]. Since the defect formation energy is much lower near the oxide surface compared to the bulk [30], the reducing $CO+CO_2$ atmosphere creates the high concentration of oxygen vacancies near the surface of the $Hf_xZr_{1-x}O_2$ nanoparticles. As a result, the oxygen-deficient $Hf_xZr_{1-x}O_2$ nanoparticles may acquire a "core-shell" type structure, in which oxygen vacancies are predominately localized within the surface layer (shell), and the central part of the particle (core) is almost free of vacancies. However, the diffraction peaks for the different o-phases could not be resolved by XRD in the $Hf_xZr_{1-x}O_2$ nanoparticles due to proximity and broadening of corresponding peaks, making it impossible to identify the *Pca21* o-phase and/or determine its fraction. Since oxygen vacancies inevitably produce Vegard strain and chemical stress [24, 31], it is likely that they can induce the desired *Pca21* o-phase when their concentration increases.

In this work, we investigate stress-induced stabilization of ferroelectricity in spherical $HfO_2$ core-shell nanoparticles. We use a thermodynamic LGD model to obtain analytical expressions for the critical radii that define a stable range of spontaneous polarization. Calculations indicate that a reentrant ferroelectric behavior results from the competition between depolarization effects and shell-induced chemical stress. This interplay provides a mechanism for ferroelectric state stabilization in nanoscale $HfO_2$ core-shell systems. We note that dense nanocomposites and nanograined ceramics consisting of FE core-shell $HfO_2$ nanoparticles with large spontaneous polarization and minimal misorientation of crystallographic axes can be prepared by tape casting or cold pressing of the core-shell $HfO_2$ nanoparticles in a large orienting electric field, or by reactive templated grain growth (see e.g., Ref. [32] and refs. therein). Such nanomaterials may overcome the principal thickness limitation of FE $HfO_2$ films (less than 20 – 30 nm [17 - 19]) and thus can be important for hafnia-based high-performance piezoelectric actuators, pyroelectric sensors, and electrocaloric devices with a minimal risk of electric breakdown.

## II. PROBLEM GEOMETRY AND MAIN ASSUMPTIONS

We consider a spherical $HfO_2$ core-shell nanoparticle with a single-domain core characterized by a spontaneous polarization $\vec{P}_3$ directed along the polar axis "3". The core is considered as defect-free, crystalline, and insulating. The core is covered with a thin shell, whose thickness $\Delta R$ is much smaller than the core radius. The core radius is $R_c$, and the outer radius is $R_s = \Delta R + R_c$. Due to the well-established fact that the defect formation energy is much lower near the oxide surface compared to the bulk [30], we assume that the reducing $CO+CO_2$ atmosphere promotes the formation of oxygen vacancies primarily in the shell rather than in the $HfO_2$ core. Therefore, the shell is assumed to be semiconducting and non-ferroelectric due to the high concentration of free ionic-electronic charges and elastic defects, primarily oxygen vacancies. We assume that the effective screening length in the shell, $\lambda_{eff}$, is smaller than the critical value $\lambda_{cr}$. For $\lambda_{eff} < \lambda_{cr}$, the free charges in the shell provide effective screening of the bound charges associated with the spontaneous polarization in the core, thereby preventing domain formation and ensuring that the single-domain approximation is self-consistent. The critical value $\lambda_{cr}$ depends on $R_c$, $W_s$, and $T$. As a rule, $\lambda_{cr}$ is about 0.1 – 0.5 nm at room temperature, $R_c \geq 5$ nm, and $\Delta R$~1 nm. For $\lambda_{eff} \geq \lambda_{cr}$, insufficient screening leads to a strong depolarization field that destabilizes the single-domain state in the core and favors the formation of a multidomain state. Under these conditions, the single-domain approximation is no longer applicable, and the finite element method (FEM) [33, 34] or the phase-field approach [35] should be used to account for possible domain formation in the nanoparticle.

We also assume that the elastic defects (e.g., oxygen vacancies) in the shell induce strong chemical strains, denoted as $w_{ij}^{s}$. The chemical strains are regarded as isotropic, $w_{ij}^{s} = \delta_{ij} w_s$, where $\delta_{ij}$ is the Kronecker-delta symbol and $w_s$ is the magnitude of the strains. These strains, which originate from, e.g., oxygen vacancies and/or elastic dipoles, are proportional to the product of the Vegard strain tensor $W_s$ and the defect concentration $n$, such that $w_s \cong W_s n$. The magnitude of $W_s$ is about (1 – 3)$\cdot 10^{-29}$ m$^3$ in oxide ferroelectrics [36, 37]. Assuming an atomic concentration $n_0 \cong 10^{29}$ m$^{-3}$, we estimated the defect concentration as $n \approx (1-5) \cdot 10^{27}$ m$^{-3}$. This value of $n$ is consistent with many experiments [38, 39, 40], which show that the defect concentration can exceed 2 – 5 % near the surface of various oxides. Hence, the range of chemical strains $w_s$ can be taken as about 1 – 5 % for the purpose of this work. We should note that Vegard strains larger than several % can induce structural relaxation, defect clustering, and/or lattice instability, which cannot be considered within the continuum LGD approach. To justify the feasibility of such large strains, we refer to the experiments for oxides [30, 38-40], as well as ab initio calculations [15, 16] corroborated by experiments [12, 13], where very large mismatch strains were applied to induce the ferroelectricity in thin $Hf_{1-x}Zr_xO_2$ films.

Due to the elastic mismatch at the core-shell interface, the chemical strains induce elastic stress in the core. In addition to defect-induced chemical strains and stresses, the contribution of the intrinsic surface stress [41] (referred to as "surface tension") associated with the particle curvature should also be considered. The intrinsic surface stress contribution is equal to $-\frac{2\mu}{R_c}$, where $\mu$ is the surface tension coefficient and its estimated value $|\mu| \cong 1-2$ N/m is typical for oxide ferroelectrics [42, 43].

The relative dielectric permittivity tensor of the shell, $\varepsilon_{ij}^{s}$, is taken to be cubic, $\varepsilon_{ij}^{s} = \delta_{ij}\varepsilon_s$, and the dielectric permittivity $\varepsilon_s$ can be relatively high due to the paraelectric state of the shell. The core-shell nanoparticle is embedded in a paraelectric (e.g., $SrTiO_3$, $KTaO_3$) or dielectric (polymer, gas, liquid, air, or vacuum) environment with a relative dielectric permittivity $\varepsilon_e$. It has been shown [44] that $\lambda_{eff}$ is related to the Debye-Hückel length $L_D$ as $\lambda_{eff} = \frac{L_D}{\varepsilon_s}$, where $L_D = \sqrt{\frac{\varepsilon_0 \varepsilon_s k_B T}{2e^2 n}}$, $\varepsilon_0$ is the vacuum dielectric constant, $k_B$ is the Boltzmann constant, $T$ is the absolute temperature, and $e$ is the elementary charge. The core-shell geometry is shown in **Fig. 1(a)** and a block-scheme of the main assumptions is shown in **Fig. 1(b)**.

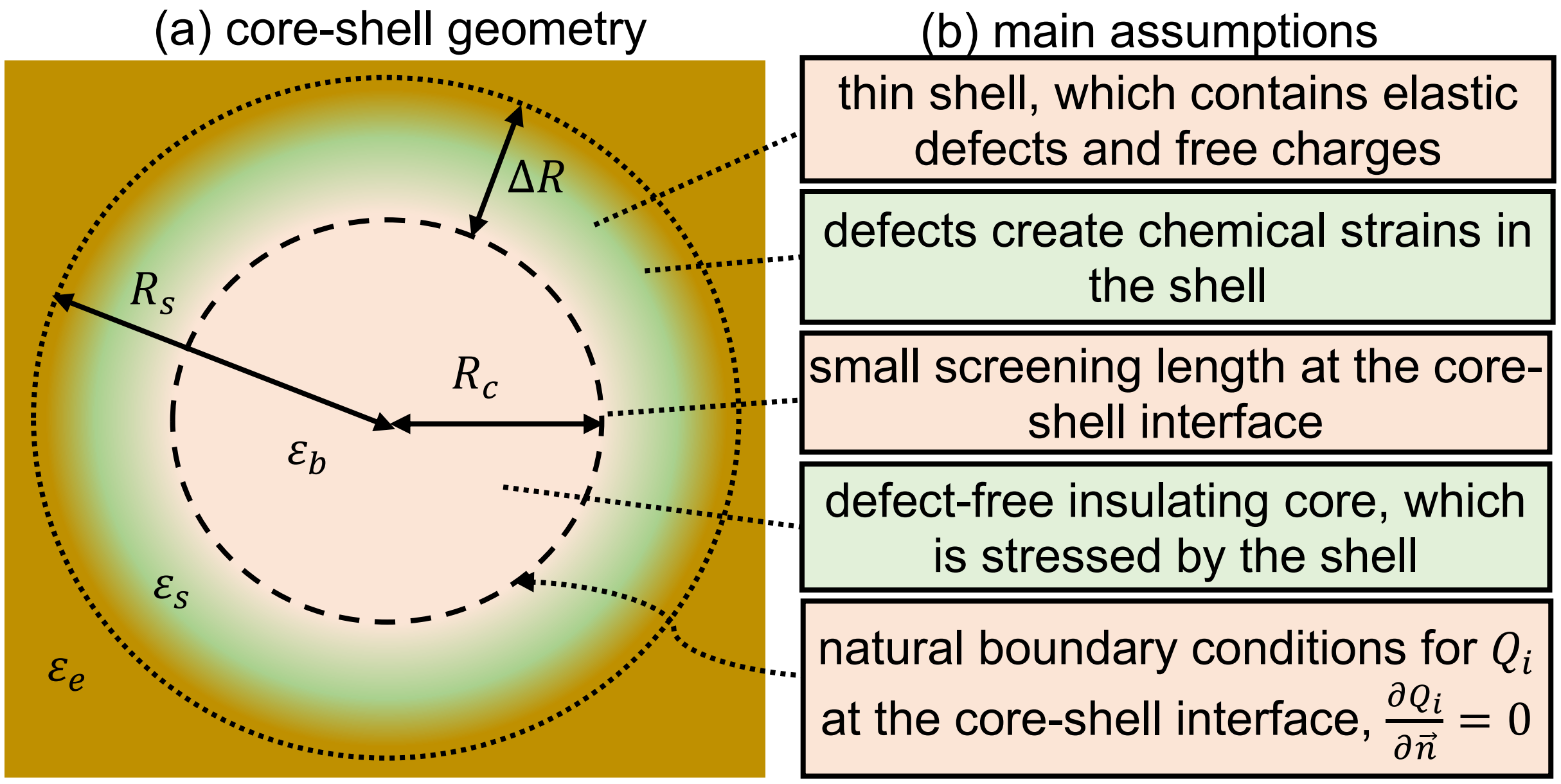

**FIGURE 1. (a)** The cross-section of a spherical $HfO_2$ core-shell nanoparticle: a defect-free core of radius $R_c$ is covered with a shell of thickness $\Delta R$, which is full of elastic defects (e.g., oxygen vacancies) and free charges. The nanoparticle is placed in an isotropic dielectric medium; $\varepsilon_b$, $\varepsilon_s$, and $\varepsilon_e$ are the core background, shell, and ambient dielectric permittivities. **(b)** Block-scheme of the main assumptions. Part (a) is adapted from Ref. [31].

## III. FREE ENERGY FUNCTIONAL OF HAFNIA CORE-SHELL NANOPARTICLES

To study the influence of chemical stresses on the phase diagrams and polar properties (e.g., spontaneous polarization and related ferroelectric characteristics) of spherical $HfO_2$ nanoparticles with a core-shell structure, we use the LGD free energy functional that includes higher-order contributions together with trilinear and biquadratic couplings involving polar, antipolar, and nonpolar order parameters [17]. The form of the free energy functional is based on results of Delodovici et al. [15, 45], and Jung and Birol [46, 47], who established the principal role of the trilinear coupling. The electrostriction coupling coefficients of $HfO_2$ were determined from the piezoelectric response reported by Datta et al. [48].

The LGD free energy of the $HfO_2$ nanoparticle core, corresponding to the FE o-phase, $F_{o-phase}$, is written as the volume integral of the free energy densities $f_{bulk}$, electric energy $f_{el}$, and surface energy $F_S$:

$$F_{o-phase} = \int (f_{bulk} + f_{el}) dV + F_S, \qquad f_{bulk} = f_{bq} + f_{tr} + f_{est} + f_{grad}. \tag{1}$$

The bulk energy density $f_{bulk}$ is the sum of biquadratic (denoted as $f_{bq}$) and trilinear (denoted as $f_{tr}$) coupling terms involving the polar, antipolar and nonpolar order parameters; elastic, electrostriction, antipolar, and nonpolar-type striction energy contributions (collectively denoted as $f_{est}$); and the gradient energy of the order parameters (denoted as $f_{grad}$). The energy $f_{bq}$ is an expansion over even 2-4-6-8 powers and $f_{tr}$ is an expansion over odd 3-5-7 powers of the dimensionless amplitudes $Q_{\Gamma 3}$, $Q_{Y2}$, and $Q_{Y4}$ of the polar phonon mode $\Gamma_{3-}$, nonpolar phonon mode $Y_{2+}$, and antipolar phonon mode $Y_{4-}$ (see Refs. [17, 45, 15, 46, 47] for details). The energies $f_{bq}$, $f_{tr}$, $f_{est}$, $f_{grad}$, and $f_{el}$ are listed in **Appendices S1-S2** in **Supplementary Materials** [49]. In a general case, the surface energy $F_S$ is the quadratic form of $Q_i$, namely $F_S = \int \alpha_i Q_i^2 dS$, where $Q_i = Q_{\Gamma 3}$, $Q_{Y2}$, and $Q_{Y4}$ represent the order parameters. Given the natural boundary conditions at the core-shell interface, $\frac{\partial Q_i}{\partial \vec{n}} = 0$, the surface energy can be assumed to be zero.

The polarization $P_3$ is proportional to the amplitude $Q_{\Gamma 3}$ of the $\Gamma_{3-}$ mode [17, 15, 46]:

$$P_3 = \frac{Z_B^* d}{V_{f.u.}} Q_{\Gamma 3} \approx P_0 Q_{\Gamma 3}, \tag{2}$$

where $Z_B^*$ is the effective Bader charge [47], $d$ is the atomic displacement corresponding to the polar mode, $V_{f.u.}$ is the formula unit (f.u.) volume, and $P_0$ is the polarization amplitude. The values $V_{f.u.} \approx 134$ Å$^3$ and $P_0 \approx 54.8$ µC/cm$^2$ at room temperature [15].

Following Refs. [17, 15], we consider the nonpolar t-phase as the reference “aristo-phase” of the $HfO_2$ core. The FE o-phase can be stable in the core when its free energy $F_{o-phase}$ is smaller than the energy of the m-phase, $F_m = \int f_m dV$, where $f_m \approx$-92 meV/f.u. is the energy density of a bulk $HfO_2$ m-phase [15]. The energy of the m-phase is significantly lower than the energy of the t-phase, which is set as zero energy. Note that the choice of zero energy is arbitrary, it is convenient to relate it with the physical mechanisms of the polarization switching in the $HfO_2$ core. Following Refs. [8-10], we assume that the polarization switching proceeds via the nonpolar t-phase; it seems reasonable to set its energy equal to zero. Thus, the critical sizes of the ferroelectricity appearance/disappearance in the $HfO_2$ core-shell nanoparticles can be estimated from the phase stability boundary condition

$$F_{o-phase} = F_m. \quad (3)$$

Because the various energy contributions exhibit different size dependencies, the phase stability condition may yield several critical core radii, $R_{cr}^{(i)}$, where $i = 1, 2$, or more. At small $R_c$, the depolarization field energy destabilizes the ferroelectric state of the core; defect-induced chemical stress induces and stabilizes the FE o-phase at intermediate sizes; and the bulk m-phase becomes energetically favorable at large $R_c$. As a result, one can expect that the ferroelectricity may exist only within a finite size window $R_{cr}^{min} < R_c < R_{cr}^{max}$. This expectation is justified by the results of calculations presented in the next section.

Elastic and/or chemical stresses, surface tension, polarization gradient, and depolarization field energies "renormalize" the coefficients in the free energy (1). Elastic stresses $\sigma_{ij}^c$, induced by the chemical strain and intrinsic surface stress, have the following form in the nanoparticle core [49]:

$$\sigma_{11}^c = \sigma_{22}^c = \sigma_{33}^c \approx \frac{-2\left(R_s^3 - R_c^3\right)\left(q_c P_0^2 Q_{\Gamma 3}^2 + z_c Q_{\Gamma 3} Q_{Y2} Q_{Y4} - w_s\right) - 3R_s^3 \left(s_{11}^s + s_{12}^s\right)\frac{2\mu}{R_c}}{2\left(R_s^3 - R_c^3\right)\left(s_{11}^c + 2s_{12}^c\right) + R_s^3\left(s_{11}^s - s_{12}^s\right) + 2R_c^3\left(s_{11}^s + 2s_{12}^s\right)}. \quad (4)$$

Here, $s_{ij}^c$ and $s_{ij}^s$ are the elastic compliances of the core and shell, respectively; $q_c = (Q_{13}^c + Q_{23}^c + Q_{33}^c)/3$ and $z_c = (Z_{13}^c + Z_{23}^c + Z_{33}^c)/3$ are the isotropic parts of the electrostriction and trilinear striction tensors of the core, respectively (see **Appendices S3-S4** [49]). The non-diagonal stresses are absent, $\sigma_{12}^c = \sigma_{13}^c = \sigma_{23}^c = 0$. Note that the contribution of the intrinsic surface stress cannot add more than 0.3 % to the stress of the nanoparticle core with radius $R_c \geq 5$ nm, since the combination of elastic compliances $s_{11}^s + 2s_{12}^s$ does not exceed $3 \cdot 10^{-12}$ $Pa^{-1}$ in the $HfO_2$ material. For the considered case $\Delta R \ll R_c$, the stress $\sigma_{ii}^c \approx \frac{-2\Delta R\left(q_c P_0^2 Q_{\Gamma 3}^2 + z_c Q_{\Gamma 3} Q_{Y2} Q_{Y4} - w_s\right)}{R_c\left(s_{11}^s + s_{12}^s\right)} - \frac{2\mu}{R_c}$.

Analytical calculations and FEM were performed to determine the applicability limits of the step-like core-shell model (4) relative to the diffuse step model and the continuous exponential decay of the defect concentration from the surface to the center of the nanoparticle (see **Appendix S5** [49] for details). We used exponential and Fermi functions to describe the defect concentration because this choice is well-grounded by experimental studies and ab initio calculations of the defect formation energy (see Ref. [30] and references therein). The obtained results confirm that the average elastic fields in the core calculated using the step-like core-shell model differ by no more than 5 % from those calculated using the diffuse step and exponential decay models, provided that the total number of defects is the same. This result has a relatively simple explanation based on the Saint-Venant principle and the direct variational method. Indeed, we consider spatially averaged quantities in a free energy formulation with renormalized coefficients instead of local distributions of elastic and electric fields (see e.g., Ref. [50]). The integration of the elastic fields over the core volume smooths out local differences between the defect distribution functions, making the result depend primarily on their average value. Therefore Eq. (4) is "robust" with respect to the details of defect distribution function and is primarily sensitive to the average defect concentration $n_d$, shell thickness $\Delta R$, and core radius $R_c$, because their product is proportional to the total number of defects $N_d$. For thin shells ($R_c \gg \Delta R$), the total number of defects, $N_d \approx \frac{4\pi}{3}(R_s^3 - R_c^3)n_d \approx 4\pi R_c^2 \Delta R n_d$, should be conserved for the validity of the core-shell model.

Note that the LGD phenomenology incorporating trilinear coupling was introduced in Ref. [17] for *epitaxial* thin $HfO_2$ films and cylindrical nanoclusters clamped to a rigid substrate. However, results of Ref. [17] are not applicable for nanoparticles. The ferroelectricity and related phenomena considered in this work depend critically on the magnitude and sign of chemical stresses created by the defect-enriched shell. The analytical form of the chemical stress in Eq. (4) differs fundamentally from the elastic strains in epitaxial thin films and/or nanoclusters with out-of-plane polarization. The physical difference is that the isotropic parts of the electrostriction and trilinear striction tensors of the core couple the emerging polarization and chemical stresses, and the stresses are radius-dependent (similarly to surface tension but in a more complex way). Anisotropic combinations of electrostriction and trilinear striction tensors, as well as the appearance of misfit dislocations, play a decisive role in the behavior of thin epitaxial films and nanoclusters clamped to a substrate.

The depolarization field $E_3^d$ and external field $E_3^e$ inside the single-domain spherical core with the ferroelectric polarization $P_3(\vec{r})$ directed along the polar axis "3" have the following form [24]:

$$E_3^d \approx -\frac{1}{\varepsilon_b + 2\varepsilon_s + (R_c/\lambda_{eff})}\frac{P_3}{\varepsilon_0}, \tag{5a}$$

$$E_3^e \approx \frac{3\varepsilon_e}{\varepsilon_b + 2\varepsilon_s + (R_c/\lambda_{eff})}E_3^0. \tag{5b}$$

Comparison of the electric fields, calculated by FEM in a self-consistent manner, with the approximate expressions in Eq. (5) showed that they are accurate to within a few percent for single-domain nanoparticles covered by a layer of screening charge with an effective screening length $\lambda_{eff}$ less than $\lambda_{cr}$. The critical value of $\lambda_{cr}$ is approximately 0.1 – 0.5 nm at room temperature [24]. For $\lambda_{eff} < \lambda_{cr}$, the free charges in the shell provide effective screening of the spontaneous polarization in the core, thereby preventing domain formation and ensuring that the single-domain approximation is self-consistent. We verified that the inequality $\lambda_{eff} < \lambda_{cr}$ was satisfied during the simulations.

In what follows, we assume that the dielectric permittivity of the shell $\varepsilon_s$ and the ambient permittivity $\varepsilon_e$ are approximately the same values, $\varepsilon_s \approx \varepsilon_e$. This assumption is justified in the case when elastic defects (e.g., oxygen vacancies) accumulate in a thin shell ($\Delta R \sim 1$ nm) at the surface of nanoparticles, as well as in the case of densely packed nanoparticles (e.g., dense ceramics), when the role of the environment is played by the particles themselves (e.g., in the effective medium approximation).

To derive a relatively simple equation for the critical radii of the nanoparticle core, we assume that the average spontaneous polarization $\bar{P}_s$ depends only weakly on stresses and core radius in the stability region of the FE o-phase (sufficiently far from the phase boundary). Using these assumptions and the strong inequality $\Delta R \ll R_c$, an approximate algebraic equation for the critical core radius, $R_{cr}$, acquires the form:

$$\frac{\lambda_{eff}}{2\varepsilon_0[(\varepsilon_b+2\varepsilon_s)R_{cr}+\lambda_{eff}]}+\left[\frac{6\Delta R}{s_{11}^s+s_{12}^s}\left(q_c\bar{P}_s^2+z_c\bar{Q}_{Y2}\bar{Q}_{Y4}\frac{\bar{P}_s}{P_0}-w_s\right)+\mu\right]\frac{2q_c}{R_{cr}}=-\Delta_\beta, \tag{6}$$

where the positive parameter $\Delta_\beta$ is estimated at the end of **Appendix S3** [49]. If we assume that the parameter $\Delta_\beta$ is size and stress independent, Eq.(6) becomes a quadratic equation. Under definite conditions it may have two positive roots, denoted earlier as $R_{cr}^{min}$ and $R_{cr}^{max}$. The analytical expressions for the critical sizes of $HfO_2$ nanoparticles can be generalized for $Hf_xZr_{1-x}O_2$ nanoparticles, provided that the corresponding free energy parameters are known from first-principles calculations.

Results presented in next section are calculated numerically using the Monte Carlo method to minimize the free energy functional (1). Equation (2) is used to recalculate the ferroelectric polarization $P_3$ from the amplitude $Q_{\Gamma 3}$ of the $\Gamma_{3-}$ mode. Condition (3) is used to calculate the phase boundary between the FE o-phase and the nonpolar m-phase. For the defect distribution, we use a diffuse step function described by a Fermi-like function $\frac{1}{1+\exp\left(\frac{R_c-r}{\delta R}\right)}$ with a finite width $\delta R \ll R_c$ (see **Appendix S5** [49] for details). It should be emphasized that Eq.(6) was not used to calculate the results presented below, because Eq.(6) loses accuracy near the phase boundaries. Equation (6) should be considered as a "scaling law" that describes the critical sizes of the $HfO_2$ core-shell nanoparticles only qualitatively. The numerical results presented in Section IV were obtained and visualized using Mathematica 14.0 [51], whose built-in Monte Carlo algorithms and related optimization routines accelerate the calculations.

## IV. RESULTS AND DISCUSSION

As argued in the introduction, the mechanisms of the FE o-phase stabilization in nanoscale $HfO_2$ remain under debate. Moreover, an interesting question arises: why is a stable FE o-phase observed in strained $HfO_2$ thin films but not in $HfO_2$ nanoparticles? An answer that follows from the calculated results and figures presented below is that relatively large compressive chemical stresses are necessary to induce the FE o-phase in the nanoparticles. Such stresses are very unlikely to arise spontaneously; instead their creation requires special treatment of the nanoparticles.

### A. Stress-Induced Effects of the Phase Diagrams, Spontaneous Polar, Antipolar and Nonpolar Order Parameters of Hafnia Core-Shell Nanoparticles

The spontaneous polarization $\bar{P}_S$ and the antipolar order parameter $\bar{Q}_{Y4s}$ of the $HfO_2$ core-shell nanoparticles as a function of defect concentration $n$ and core radius $R_c$ are shown in **Fig. 2(a)-2(f)** for several values of the ambient medium dielectric permittivity ($\varepsilon_e$): $\varepsilon_e = 3$ (corresponding to a low-k medium such as oxides, polymers poly-vinyl-butyral and ethyl-cellulose), $\varepsilon_e = 30$ (corresponding to a high-k polymer such as PV(DF-TrFE)), and $\varepsilon_e = 300$ (corresponding to a paraelectric $SrTiO_3$). Thus, the choice of $\varepsilon_e$ is linked to real materials, which are used as matrices for the preparation of ferroelectric nanocomposites (see e.g., Refs. [52, 53, 54] and references therein).

In **Fig. 2**, the shell thickness is fixed at $\Delta R$ =1.0 nm, the Vegard tensor magnitude is set to $W_s = -1 \cdot 10^{-29}$ m$^3$, and the temperature is $T = 300$ K. In the calculations, the effective screening length is assumed to depend on the defect concentration $n$ as $\lambda_{eff} = \sqrt{\frac{\varepsilon_0 k_B T}{2e^2 n \varepsilon_s}}$, and the condition $\varepsilon_s \cong \varepsilon_e$ is imposed. The expression for $\lambda_{eff}$ is derived in Ref. [44] using the Debye-Hückel approximation. The dependence of the chemical strain $w_s$ on the defect concentration $n$ obeys the linear Vegard law, $w_s \cong W_s n$.

The FE o-phase is stable only under compressive chemical strain $w_s \leq -(0.7 - 2.5)\%$, corresponding to $W_s < 0$ and $n > n_{min}$. The requirement of applying compressive strain to stabilize the FE o-phase in the $HfO_2$ core-shell nanoparticles is consistent with DFT results reported for thin films [14, 17, 15]. The region of the nonpolar m-phase stability decreases significantly as $\varepsilon_e$ increases, which is replaced by the FE o-phase (compare the first, second, and third columns in **Fig. 2**). The FE o-phase is reentrant for $\varepsilon_e \leq \varepsilon_{max}$ and defect concentrations $n > n_{min}$. The maximal value $\varepsilon_{max} \geq 300$ for material parameters of $HfO_2$, $W_s$ =–$1\cdot10^{-29}$ m$^3$, and room temperature. The minimal concentration of defects, $n_{min}$, decreases with an increase in $\varepsilon_e$ from $2.5\cdot10^{27}$ m$^{-3}$ at $\varepsilon_e = 3$ to $7\cdot10^{26}$ m$^{-3}$ at $\varepsilon_e = 300$ (see the black circles in **Fig. 2(a)**, **2(b)**, and **2(c)**). Since the depolarization field scales as $E_3^d \sim \frac{\lambda_{eff}}{(\varepsilon_b + 2\varepsilon_e)R_c + \lambda_{eff}}$ according to Eq.(4), and the effective screening length scales as $\lambda_{eff} \sim \sqrt{\frac{1}{\varepsilon_e}}$, both $E_3^d$ and $\lambda_{eff}$ decrease simultaneously with an increase in $\varepsilon_e$. Thus, the simultaneous reduction of the m-phase stability region, the minimal concentration $n_{min}$, and the depolarization field $E_3^d$ can be attributed to enhanced dielectric and space-charge screening with increasing $\varepsilon_e$.

Note that the values of $\lambda_{eff}$ corresponding to the region of FE o-phase stability in **Fig. 2** are significantly smaller than the critical value $\lambda_{cr}$ required for domain formation. The FEM, which accounts for possible domain formation, shows that the FE o-phase disappears completely for $\lambda_{eff} > \lambda_{max}$, where $\lambda_{max} > \lambda_{cr}$ and increases with increasing magnitude of $W_s$. The calculations further indicate that domain formation slightly reduces the critical radius for ferroelectricity disappearance. A detailed analysis of domain formation in the $HfO_2$ core-shell nanoparticles will be presented elsewhere.

The color scale in **Fig. 2(a)-2(c)** represents the absolute value of the spontaneous polarization $\bar{P}_s$ in the deepest potential well (i.e., deepest minimum) of the free energy given by Eq.(1). The spontaneous polarization can reach relatively large values ($\bar{P}_s$ > 60 – 70 μC/cm$^2$) for small core radii ($R_c$ < 10 nm) and high elastic defect concentrations ($n$ >10$^{27}$ m$^{-3}$). Although the required values of $n$ are high, they are quite realistic and decrease strongly (from 5·10$^{27}$ m$^{-3}$ to 5·10$^{27}$ m$^{-3}$) with an increase in $\varepsilon_e$ from 3 to 300. The spontaneous polarization of the $HfO_2$ core exceeds the experimentally observed values in $Hf_xZr_{1-x}O_2$ thin films by a factor of 3 [7]. The increase arises due to compressive chemical stress in the shell.

The color scale in **Fig. 2(d)-2(f)** represents the absolute value of the antipolar mode $\bar{Q}_{Y4s}$ in the deepest potential well of the free energy (1). The sharp boundary between the FE o-phase (with $\bar{P}_s > 0$ and $\bar{Q}_{Y4s} > 0$) and nonpolar m-phase (with $\bar{P}_s = 0$ and $\bar{Q}_{Y4s} = 0$) corresponds to the first-order phase transition curve defining the critical radius $R_{cr}$ as a function of defect concentration $n$.

It is seen from **Fig. 2(g)-2(i)**, that the difference between the antipolar and nonpolar order parameters amplitudes, $|\bar{Q}_{Y4s}| - |\bar{Q}_{Y2s}|$, is very small in the free energy minimum, reflecting the negligible anisotropy of the deepest potential well with respect to the $Y_{4-}$ and $Y_{2+}$ modes [17]. Specifically, the difference $|\bar{Q}_{Y4s}| - |\bar{Q}_{Y2s}|$ changes its sign along the curved stripe in middle of the FE o-phase region and does not exceed ±0.75 pm, which is more than 30 times smaller than the maximal value of $\bar{Q}_{Y4s} \approx 0.25$ Å. The small splitting between the amplitudes of antipolar and nonpolar mode means that distortion of the o-phase is strongly mixed between these two modes (assuming angular isotropy within the framework of the model used). This result explicitly indicates that the anisotropy in the $\{Y_{4-}, Y_{2+}\}$ subspace is weak and thus the pure antipolar ($\bar{Q}_{Y4s} \neq 0$ and $\bar{Q}_{Y2s} = 0$) and nonpolar ($\bar{Q}_{Y4s} = 0$ and $\bar{Q}_{Y2s} \neq 0$) states are unstable in the o-phase.

As follows from Eq.(5), the stress-dependent upper critical core radius $R_{cr}^{max}$, corresponding to the appearance of ferroelectricity, is primarily governed by the size dependence of the chemical stress $\sigma_{ij}^{c}$, which scales as $\frac{1}{R_{cr}}$ according to Eq.(3). In contrast, the lower critical radius $R_{cr}^{min}$, corresponding to the disappearance of ferroelectricity, is primarily governed by the size dependence of the depolarization field $E_3^d$, which is proportional to the function $\frac{\lambda_{eff}}{(\varepsilon_b+2\varepsilon_s)R_{cr}+\lambda_{eff}}$ according to Eq.(4). The distinct radius scaling of chemical stress and depolarization field energy therefore produces a finite interval of core radii in which the ferroelectric phase is thermodynamically stable. For small radii, the depolarization field energy contribution dominates and suppresses the ferroelectric order, whereas at large radii the chemical stress-induced stabilization weakens and restores the nonpolar m-phase.

The critical radii $R_{cr}^{max}$ and $R_{cr}^{min}$ depend on the defect concentration $n$ and the dielectric permittivity of the surrounding medium $\varepsilon_e$ (see the black circles in **Fig. 2(d)**, **2(e)**, and **2(f)**). Both radii, $R_{cr}^{min}$ and $R_{cr}^{max}$, exist only for $n > n_{min}$; they merge at $n \to n_{min}$ and disappear for $n < n_{min}$. For a broad range of $\varepsilon_e$, namely, $1 \leq \varepsilon_e \leq \varepsilon_{max}$, the FE o-phase is thermodynamically stable within the finite window $R_{cr}^{min} < R_c < R_{cr}^{max}$. For large permittivities, $\varepsilon_e > \varepsilon_{max}$, the lower critical radius decreases below the limiting size of the continuous LGD theory applicability. The limit is determined by several correlation lengths, which is about 5 – 10 lattice constants. Thus, in this regime, $R_{cr}^{min}$ is limited by several correlation lengths, because the depolarization field energy becomes negligibly small due to the effective dielectric screening at $\varepsilon_e > \varepsilon_{max}$.

The deepest minimum of the free energy density, $f_{o-phase}^{min}$, as a function of defect concentration $n$ and core radius $R_c$, calculated for $\varepsilon_e = 3$, 30, and 300, is shown in **Fig. 2(g), 2(k)**, and **2(l)**, respectively. The red background in the figure corresponds to the bulk m-phase with the energy density $f_m = -92$ meV/f.u. [15]. The first-order phase transition between the o-phase and m-phase is determined by the condition $f_{o-phase}^{min} = f_m$. The phase boundary is shown by the black curves in **Fig. 2(g)-2(l)**. As expected, the deepest minimum in the o-phase, $f_{o-phase}^{min}$, occurs at large defect concentrations and small core radii (see the upper left region of the diagrams in **Fig. 2(g)-2(l)**), where defect-induced compressive chemical stress energetically favors the o-phase over the m-phase, driving the ferroelectric phase stabilization.

It should be emphasized that the continuum LGD description is quantitatively valid for system sizes larger 10 lattice constants, which corresponds to 5 nm diameter $HfO_2$ nanoparticles (see the dashed vertical lines in **Fig. 2**). The 5-nm nanoparticle contains about 490 f.u., which significantly exceeds the correlation volume (~20 f.u.) required for the formation of long-range order. Results shown in **Fig. 2 (**as well as in **Fig. 3**) for smaller sizes (e.g., for $R_c < 2.5$ nm) may be only qualitatively valid. For larger radii, the applicability of the LGD theory is corroborated by the fact that the critical sizes for the onset of the long-range order calculated from atomistic and phenomenological theories are in good agreement with each other, as well as with experimental results for other ferroelectric systems [55, 56, 57, 58]. Important that the role of thermal fluctuations is negligible in ferroelectric nanoparticles over a wide range of sizes, including the immediate vicinity of the size-induced transition to the ferroelectric phase, because the height of the potential barrier associated with the onset of spontaneous polarization greatly exceeds the thermal energy (see, e.g., **Fig. 2(b)** in Ref. [59]). In the considered case of a 5-nm $HfO_2$ nanoparticle, the barrier is more than $10^2 k_B T$ at room temperature, since the transition from the m-phase to the FE o-phase is of the first order. The contrasts with nanosized ferromagnets, where the potential barriers are typically shallow.

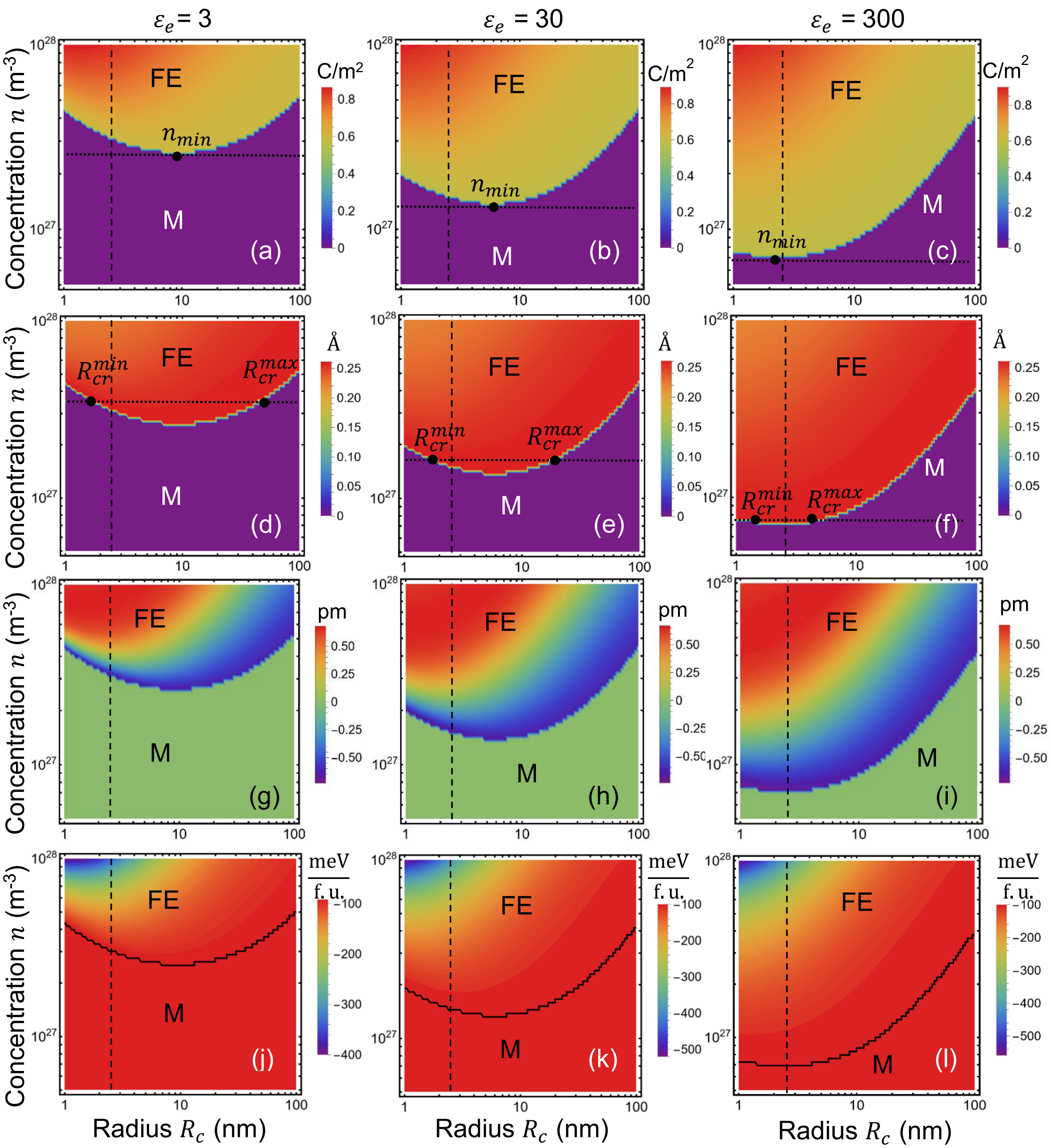

**FIGURE 2.** The absolute value of the spontaneous polarization $\bar{P}_S$ **(a, b, c)**, the amplitude of the antipolar order parameter $|\bar{Q}_{Y4s}|$ **(d, e, f)**, the difference $|\bar{Q}_{Y4s}| - |\bar{Q}_{Y2s}|$ **(g, h, i)**, and the deepest minimum of the free energy density $f_{o-phase}^{min}$ **(j, k, l)** as a function of core radius $R_c$ and defect concentration $n$ calculated for several values of the ambient permittivity $\varepsilon_e = 3$ **(a, d, g, j)**, 30 **(b, e, h, k)**, and 300 **(c, f, i, l)**. Abbreviation "FE" is the FE o-phase, "M" denotes the nonpolar m-phase of the $HfO_2$ nanoparticle. Dashed vertical lines indicate the size limit for the applicability of the LGD approach. $\Delta R = 1$ nm, $W_s = -1 \cdot 10^{-29}$ m$^3$, $\mu = 0$, $V_{f.u.} \approx 134$ Å$^3$, $T = 300$ K, and $\varepsilon_s \cong \varepsilon_e$. Material parameters of $HfO_2$ used in calculations are listed in **Tables S1-S2** in **Appendix S1**

of Ref. [17].

### B. Polarization Switching, Dielectric and Piezoelectric Properties of Hafnia Core-Shell Nanoparticles: from Modeling to Applications

Note that the same LGD free energy functional that determines the chemical stress-induced stabilization of the FE o-phase also defines the switching pathway of the spontaneous polarization and the dielectric and piezoelectric properties of the $HfO_2$ core-shell nanoparticles. Below, we study the influence of chemical stress and size effects on polarization switching and polar properties of the nanoparticles.

Similarly to the case of strained $HfO_2$ thin films considered in Ref. [17], the switching path between the $-\bar{P}_s$ and $+\bar{P}_s$ states is mediated by the virtual *Ccce* phase [45, 15] in the $HfO_2$ core-shell nanoparticles. As a result, the lowest polarization switching barrier $b_{af}$ is approximately +48 meV/f.u., corresponding to the energy difference between the *Ccce* phase and the t-phase. The activation field for polarization switching can be estimated as $E_{af} \cong b_{af}/\bar{P}_s$, where $\bar{P}_s$ depends on the chemical strains $w_s$, defect concentration $n$, core radius $R_c$, and shell thickness $\Delta R$. Color maps of $E_{af}$ as a function of $R_c$ and $n$ are shown in **Fig. 3(a) – 3(c)**. The activation fields are calculated for several values of the ambient permittivity, $\varepsilon_e$ =3, 30, and 300. The values of $E_{af}$ range from 0.65 MV/cm (far from the o-m phase boundary) to 0.95 MV/cm (nearby the o-m phase boundary), and are very close to the activation fields calculated for strained $HfO_2$ thin films [17]. The lowest values of $E_{af}$ occur at large defect concentrations and small core radii, where the compressive chemical stress $\sigma_{ii}^c$ is largest. The region expands with increasing $\varepsilon_e$ due to a reduction in the depolarization field. The calculated activation fields are somewhat lower than the coercive fields $E_c \cong$ 1.05 – 1.35 MV/cm observed experimentally in 10-nm thick $Hf_xZr_{1-x}O_2$ films [7, 60]. This result is expected from the activation rate theory, because the local nucleation of nanodomains (arising at $E_3^e \approx E_{af}$) precedes the global polarization switching (occurring at $E_3^e \approx E_c$). The calculated activation fields are of the same order of magnitude as experimentally reported coercive fields, indicating that the predicted switching barriers are physically consistent with experimental values.

Color maps of the linear relative dielectric permittivity $\varepsilon_{33}$ as a function of $R_c$ and $n$ are shown in **Fig. 3(d) – 3(f)**, which are calculated for $\varepsilon_e$ =3, 30, and 300 and room temperature. As can be seen from the figure, the magnitude of $\varepsilon_{33}$ can be tuned by the varying the core radius and chemical strains from 10 to 35 at room temperature. The permittivity reaches maximal values (slightly above 30) relatively far from the boundary between the o- and m-phases. The permittivity does not diverge at the o-m boundary because the transition is of the first order. The calculated values of $\varepsilon_{33}$ ≈10 – 30 agree well with the values of effective permittivity measured at 100 – 500 kHz in the densely pressed oxygen-deficient $Hf_xZr_{1-x}O_{2-y}$ nanopowders (x = 1, 0.6, 0.5, and 0.4) with particle sizes of 7 – 10 nm [27]. Note that the nanoparticles studied in Ref. [27] were annealed in a $CO+CO_2$ atmosphere to induce the maximal amount of oxygen vacancies near the surface. Thus, calculated values of $\varepsilon_{33}$ are of the same order of magnitude as experimentally measured permittivity, being physically consistent with experimental results [27].

The region of maximal $\varepsilon_{33}$, which has the shape of a wide curved stripe located relatively far from the boundary between the o- and m-phases, can be explained by the significant contribution of the nonpolar and antipolar orders to the permittivity due to the trilinear coupling in the free energy (1). Otherwise, the permittivity should be maximal at the o-m boundary, which is not the case shown in **Fig. 3(d) – 3(f)**. As seen in the color maps of the difference between antipolar and nonpolar order parameters shown in **Fig. 2(g) – 2(i)**, the difference $|\bar{Q}_{Y4s}| - |\bar{Q}_{Y2s}|$ is minimal in the same region where $\varepsilon_{33}$ is maximal. Additional calculations show that the maximum of $\varepsilon_{33}$ occurs at $|\bar{Q}_{Y4s}| = |\bar{Q}_{Y2s}|$, as anticipated from the specific form of the free energy (1) in the 3D order-parameter space $\{\bar{P}_s, \bar{Q}_{Y4s}, \bar{Q}_{Y2s}\}$.

It should be expected that the piezoelectric response reaches maximal values far from the o-m boundary because of the piezoelectric coefficients $d_{333} \sim \varepsilon_{33}\bar{P}_s$. Color maps of the piezoelectric coefficient $d_{333}$ as a function of $R_c$ and $n$ are shown in **Fig. 3(g) – 3(i)**. The figures are calculated for $\varepsilon_e$ =3, 30, and 300 and room temperature. For all $\varepsilon_e$, the region of maximal $d_{333}$ has the shape of a wide curved stripe located relatively far from the boundary between the o- and m-phases. As should be expected, $d_{333} = 0$ in the m-phase. Calculated values of $d_{333}$ are negative in the FE o-phase and reach $-(2-3)$ pm/V, which is in a good agreement with the experimental results and *ab initio* calculations for thin $HfO_2$ layers [48]. As can be seen from **Fig. 3(g) – 3(i)**, the magnitude of $d_{333}$ can be tuned by size variation and chemical strains from nearly zero to $-3.2$ pm/V.

Note that all $d_{3ij}$ appeared negative in the FE o-phase of $HfO_2$ [48], meaning that the $HfO_2$ core-shell nanoparticles belong to so-called "auxetic piezoelectrics" [61, 62]. Auxetic piezoelectrics were introduced in Ref. [61] by analogy with auxetic materials with a negative Poisson's ratio, which expand in all directions when stretched in one direction. In "conventional" piezoelectrics, the piezoelectric coefficients $d_{322}$ and $d_{311}$ are opposite in sign to $d_{333}$, and, as a rule, $d_{322} + d_{311} \approx -d_{333}$. Therefore, their volumetric piezoelectric effect, defined by the sum of piezoelectric coefficients, $d_{VP} = d_{333} + d_{322} + d_{311}$, is an order of magnitude smaller than the separate values of piezoelectric coefficients (as a rule, $d_{VP}$ is less than $5-10$ pm/V) [61]. Unlike conventional piezoelectrics, auxetic piezoelectrics with all positive (or all negative) $d_{3ij}$ could have increased volumetric piezo-response, which is especially important for highly sensitive piezoelectric applications. As can be seen from **Fig. 3(j) – 3(l)**, the magnitude of $d_{VP}$ reaches 10 – 12 pm/V in the region of maximal $\varepsilon_{33}$. The increase of $d_{VP}$ occurs due to the joint action of trilinear coupling between the polar, antipolar and nonpolar modes, size effects, and chemical strains.

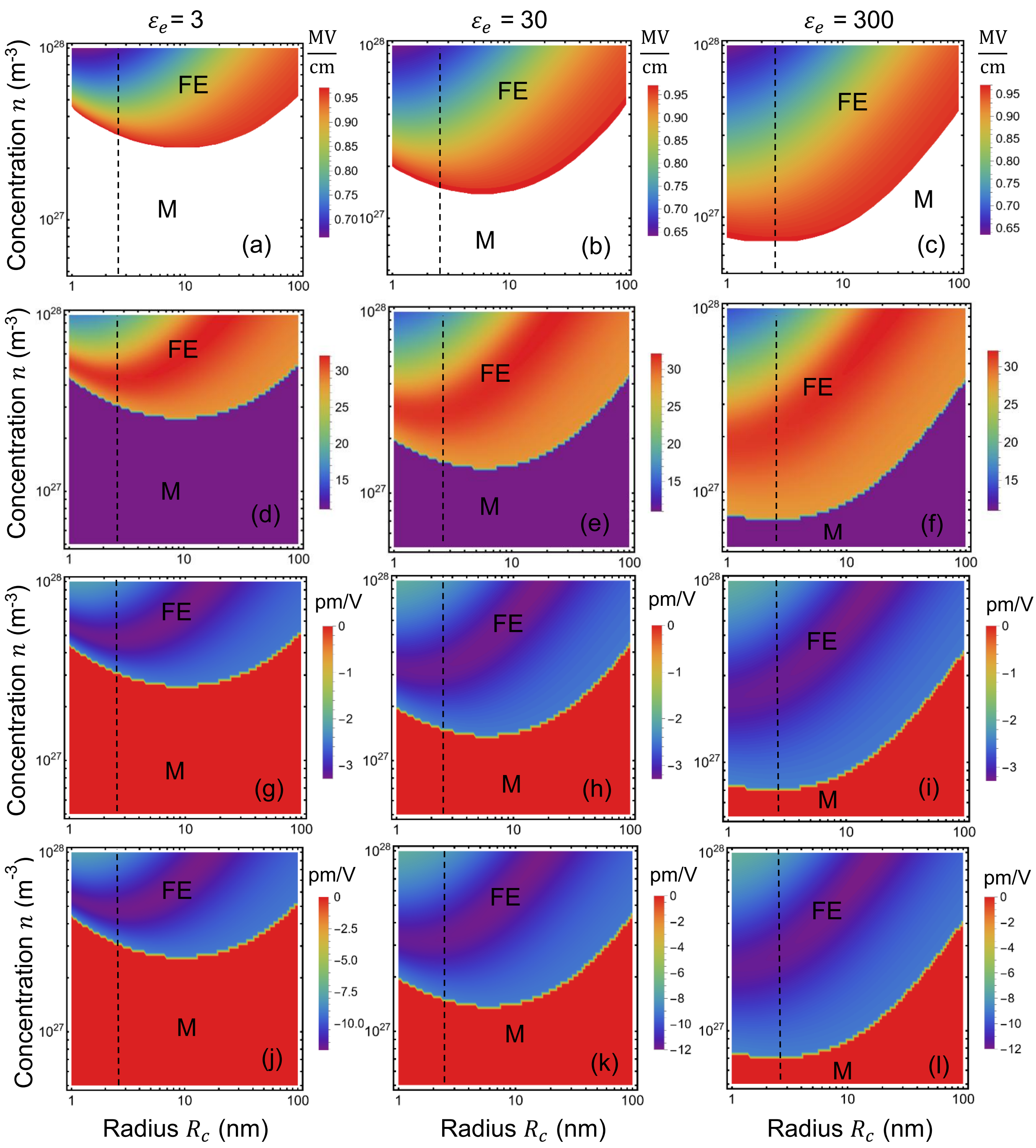

**FIGURE 3.** The activation field $E_{af}$ **(a, b, c)**, relative dielectric permittivity $\varepsilon_{33}$ **(d, e, f)**, piezoelectric coefficient $d_{333}$ **(g, h, i)** and volumetric piezo-response $d_{VP}$ **(j, k, l)** as a function of core radius $R_c$ and defect

concentration $n$ calculated for several values of the ambient permittivity $\varepsilon_e = 3$ **(a, d, g, j)**, 30 **(b, e, h, )**, and 300 **(c, f, i)**. The abbreviation "FE" is the FE o-phase, and "M" denotes the nonpolar m-phase of the $HfO_2$ nanoparticle. Dashed vertical lines indicate the size limit for the applicability of the LGD approach. $\Delta R = 1$ nm, $W_S = -1\cdot10^{-29}$ m$^3$, $\mu = 0$, $V_{f.u.} \approx 134$ Å$^3$, $T = 300$ K, and $\varepsilon_s \cong \varepsilon_e$. Material parameters of $HfO_2$ used in calculations are listed in **Tables S1-S2** in **Appendix S1** of Ref. [17].

Thus, we predict that defect engineering can induce and control the polar, dielectric, and piezoelectric properties of $HfO_2$ core-shell nanoparticles. However, improving the dielectric and/or volumetric piezoelectric response of the $HfO_2$ nanomaterials remains an important challenge for advanced electronic applications.

This result seems rather important for applications, because densely packed core-shell $HfO_2$ nanoparticles, with preferentially orientated polar axes (or perfectly aligned in the ideal case), may overcome one of the principal limitations of ferroelectric $HfO_2$ films (i.e., their thickness, typically limited to less than 20 – 30 nm [17 – 19]). The high degree of orientation of the crystallographic axes can be achieved in dense composites (or ceramics) prepared by tape casting (or cold pressing) in a large orienting electric field, as well as by reactive templated grain growth [32]. The composite layer consisting of densely packed core-shell $HfO_2$ nanoparticles with minimal misorientation of the polar axes, or a dense ceramics consisting of polarized nanosized $HfO_2$ grains, would maintain piezoelectric properties for much larger layer thicknesses (up to hundreds of microns) compared with thin epitaxial $HfO_2$ films (less than 30 nm). Increasing the piezoelectric layer thickness is required to minimize the risk of electric breakdown in high-performance piezoelectric actuators, pyroelectric sensors, and electrocaloric devices. The schematic illustration of a piezoelectric capacitor of thickness $h_f$, consisting of densely packed polarized $HfO_2$ nanoparticles placed in a high-k dielectric or paraelectric matrix (e.g., PV(DF-TrFE), rutile $TiO_2$, or $SrTiO_3$), is shown in **Fig. 4(a)**. The piezoelectric displacement of the capacitor surface is $h_{VP} \cong d_{VP}U$, where $U$ is the applied voltage (see **Fig. 4(b)**). Because the composite layer can be much thicker, substantially higher voltages can be applied while maintaining a similar electric field, allowing much larger $h_{VP}$ to be achieved than in thin $HfO_2$ films.

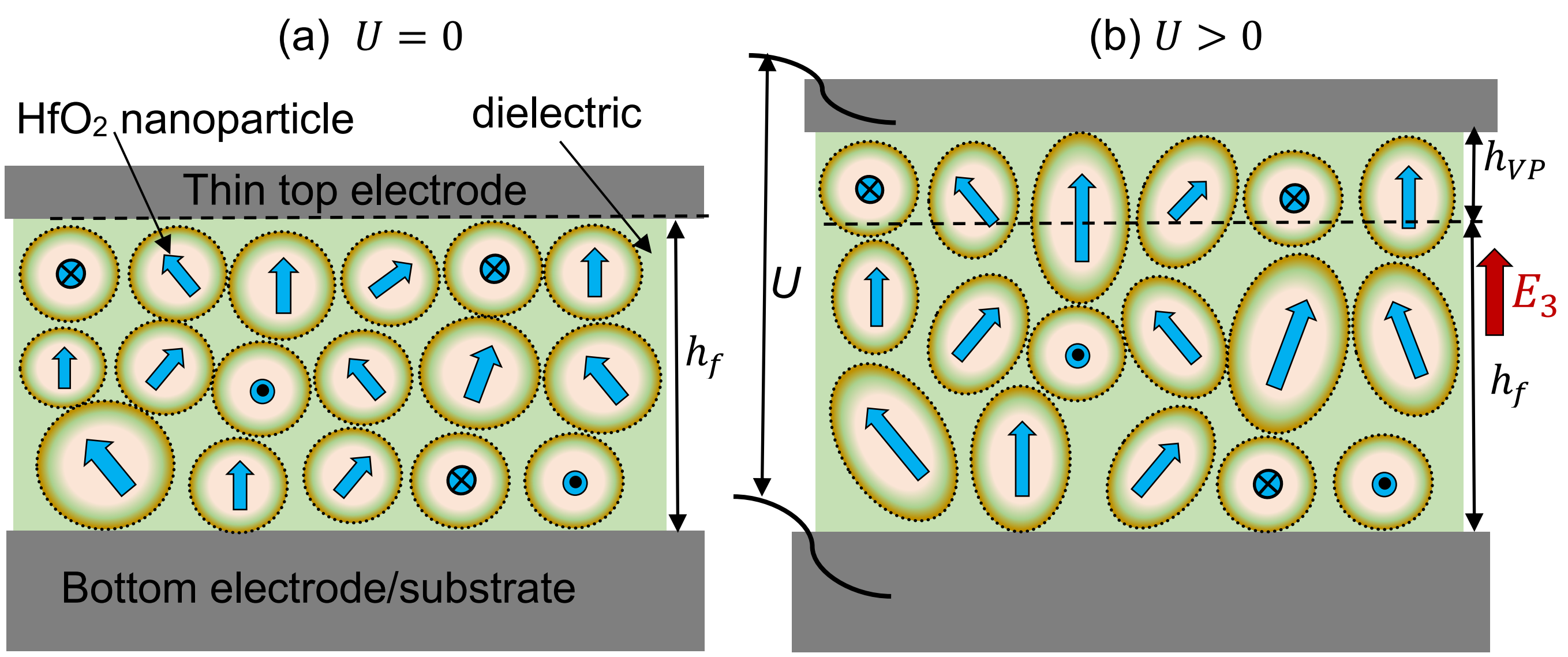

**FIGURE 4.** Schematic illustration of a piezoelectric capacitor of thickness $h_f$ consisting of densely packed polarized $HfO_2$ nanoparticles, without an applied voltage **(a)** and with an applied voltage, $U$ **(b)**. The blue arrows show the direction of the spontaneous polarization (aligned with the crystallographic direction of the polar axis) in the nanoparticle cores; the red arrow shows the direction of the electric field, $E_3 \cong \frac{U}{h_f}$.

## V. CONCLUSIONS

We have investigated the influence of chemical stresses on phase diagrams and polar properties of spherical $HfO_2$ core-shell nanoparticles using the Landau-Ginzburg-Devonshire free energy functional with higher powers, and trilinear and biquadratic couplings of polar, antipolar and nonpolar order parameters. We have shown that the reentrant stabilization of the FE o-phase arises from the competition between the size-dependent depolarization field energy, which suppresses ferroelectric order at small radii, and the stress-induced lowering of the orthorhombic free energy caused by defect-induced chemical strains in the shell.

We predict that the FE o-phase can be stable at compressive chemical strains exceeding 1 – 5 % in the shell. The minimal concentration of elastic defects, which create chemical strains, is determined by the ambient dielectric permittivity $\varepsilon_e$ and Vegard stress tensor. For ambient dielectric permittivities $\varepsilon_e \leq \varepsilon_{max}$, the FE o-phase exhibits reentrant behavior with respect to the size of nanoparticles, such that the spontaneous polarization exists only in a finite interval of core radii $R_{cr}^{min} < R_c < R_{cr}^{max}$. The lower critical radius $R_{cr}^{min}$ is governed by depolarization and correlation effects, whereas the upper critical radius $R_{cr}^{max}$ is governed by size-dependent chemical stresses in the shell. For $\varepsilon_e \geq \varepsilon_{max}$, the FE o-phase is thermodynamically stable at $R_c < R_{cr}^{max}$ and $R_{cr}^{min}$ is limited by the correlation thickness being equal to several lattice constants. Increasing the ambient dielectric permittivity decreases the energy of depolarization field and expands the ferroelectric stability window. Embedding $HfO_2$ nanoparticles in high-permittivity environments therefore provides an additional thermodynamic pathway for stabilizing ferroelectricity. The maximal value $\varepsilon_{max} \cong 300$ was calculated at room temperature, which corresponds to the $SrTiO_3$ environment of $HfO_2$ cores.

We predicted that the spontaneous polarization could reach relatively large values (>70 μC/cm$^2$) for small core radii ($R_c < 10$ nm) and realistic concentrations of elastic defects ($n >\cdot 10^{27}$ m$^{-3}$), which induce compressive chemical stress in the shell. The increase in spontaneous polarization of the $HfO_2$ core is more than 3 times compared to the values observed experimentally in $Hf_xZr_{1-x}O_2$ thin films. The predicted switching barriers within the ferroelectric stability window of the $HfO_2$ core-shell nanoparticles are of the same order of magnitude as experimentally reported coercive fields in $HfO_2$ thin films, indicating that the stabilized FE o-phase remains switchable under accessible electric fields. Thus, defect engineering opens a remarkable possibility of significant enhancement of polar properties of $HfO_2$ core-shell nanoparticles for their applications in advanced memory cells and logic devices.

We also predict that defect engineering can induce and/or control the polar, dielectric, and piezoelectric properties of $HfO_2$ core-shell nanoparticles. This prediction can be useful for elaboration of dense ceramics and composites, consisting of ferroelectric $HfO_2$ core-shell nanoparticles. The potential advantage of such nanomaterials lies in their silicon compatibility and their ability to exhibit a relatively large auxetic-type volumetric piezoelectric response (above 10 pm/V) over a wide range of external parameters, including particle size and defect concentration, without requiring operation close to the ferroelectric-paraelectric phase boundary. This is especially important for highly sensitive piezoelectric sensors, actuators, and surface acoustic wave filters.

While nominally strain-free $Hf_xZr_{1-x}O_2$ thin films can exhibit ferroelectricity [18, 19], the experimentally reported polarization values are significantly smaller than those predicted in the present work for $HfO_2$ core–shell nanoparticles (more than $50 - 70$ μC/cm$^2$), where chemical stress stabilizes the FE o-phase. This contrast suggests that, although ferroelectricity in core-shell $HfO_2$ nanoparticles is not an intrinsic bulk property, it can emerge over a finite range of particle sizes through the control of stress and screening, resulting in a comparatively large polarization. This result seems important for applications, because densely packed core-shell $HfO_2$ nanoparticles with preferentially aligned crystallographic axes, for example, in dense composites (or ceramics) prepared by tape casting (or cold pressing) in a large orienting electric field or by using the reactive templated grain growth [32], may overcome one of the principal limitations of ferroelectric $HfO_2$ films, namely their limited thickness (typically less than 20 – 30 nm [17 – 19]). Dense composite films containing polarized core-shell $HfO_2$ nanoparticles or dense ceramics with polarized nanosized grains would maintain ferroelectric properties at much larger thicknesses (up to hundreds of microns), which is required for high-performance piezoelectric actuators, pyroelectric sensors, and electrocaloric devices while minimizing the risk of electric breakdown.

Analytical expressions derived for the critical radii can be generalized for $Hf_xZr_{1-x}O_2$ nanoparticles, providing that corresponding parameters of the free energy are known from the first-principles calculations. Thus, this work identifies a stress-driven mechanism for reentrant ferroelectric phase stabilization in nanoscale $HfO_2$, arising from the competition between the depolarization field suppression and shell-induced chemical stress.

Also, obtained results may help answer the question of why the stable FE o-phase has been observed in $Hf_xZr_{1-x}O_2$ thin films but has not yet been observed in $HfO_2$ nanoparticles. We predicted that relatively large compressive chemical strains are necessary to induce the FE o-phase in the $HfO_2$ nanoparticles. However, the creation of such stress requires controllable treatment of the nanoparticles in a special environment and/or other special conditions of their synthesis. Assuming that the technological issues can be overcome, for example, by annealing $HfO_2$ nanoparticles in a $CO+CO_2$ atmosphere [26, 27, 54], the appearance of ferroelectric hysteresis and/or a domain structure in the small (≤10 nm) core-shell $HfO_2$ nanoparticles can be verified directly by switching spectroscopy piezoelectric force microscopy (SS-PFM) with dual-frequency resonance tracking (DART) [63]. For this purpose, $HfO_2$ nanoparticles should be deposited as a layer onto a conducting substrate (e.g., pyrolytic graphite). The presence of local piezoelectric response hysteresis loops and/or the direct visualization of domain structures has been demonstrated by SS-PFM and DART-PFM in nanoislands and nanoparticles of other ferroelectric materials [63]. Other experimental signatures of ferroelectricity in $HfO_2$ nanoparticles include pronounced pyroelectric and/or electrocaloric properties of pre-poled nanopowders or nanograined ceramics prepared by cold-pressing the nanoparticles.

**Acknowledgements.** The work of A.N.M. is funded by the National Research Foundation of Ukraine (project “Manyfold-degenerated metastable states of spontaneous polarization in nanoferroics: theory, experiment, and perspectives for digital nanoelectronics”, grant N 2023.03/0132). The work of E.A.E. is funded by the National Research Foundation of Ukraine (project “Silicon-compatible ferroelectric nanocomposites for electronics and sensors”, grant N 2023.03/0127). The work of S.V.K. is supported by S.V.K. start-up funds. A.N.M. also acknowledges (discussion of the results presented in the work) the Horizon Europe Framework Programme (HORIZON-TMA-MSCA-SE), project № 101131229, Piezoelectricity in 2D-materials: materials, modeling, and applications (PIEZO-2D) and the EOARD project 9IOE063 (related STCU partner project is P751c).

**Authors' contribution.** A.N.M. generated the idea of research, formulated the physical problem, performed analytical calculations of the critical size, analyzed results and wrote the manuscript draft. E.A.E. performed symmetry analysis, numerical calculations and prepared corresponding figures. S.V.K. and D.R.E. worked on improvement of the manuscript. All co-authors discussed the results.

# Supplementary Materials to the manuscript

## APPENDIX S1. LGD-type free energy derived from the DFT calculations

The compact form of the LGD free energy bulk density $f_{bulk}$ is the sum of the 2-4-6-8 powers (energy $f_{bl}$) and 1-5-7 powers (energy $f_{tr}$)of the three phonon modes, elastic and striction energy contributions (energy $f_{est}$), and the gradient energy of the order parameters ($f_{grad}$) [64]:

$$f_{bulk} = f_{bq} + f_{tr} + f_{est} + f_{grad}, \tag{S1.1a}$$

$$f_{bl} = \beta_i Q_i^2 + \delta_{ij} Q_i^2 Q_j^2 + \eta_{ijk} Q_i^2 Q_j^2 Q_k^2 + \xi_{ijkl} Q_i^2 Q_j^2 Q_k^2 Q_l^2, \tag{S1.1b}$$

$$f_{tr} = \left(\gamma + \epsilon_i Q_i^2 + \zeta_{ij} Q_i^2 Q_j^2\right) Q_{\Gamma 3} Q_{Y2} Q_{Y4}, \tag{S1.1c}$$

$$f_{est} = \frac{1}{2} c_{ijkl} u_{ij} u_{kl} - \tilde{q}_{ijkl} u_{ij} Q_k Q_l - \left(1 + v_i Q_i^2 + k_{ij} Q_i^2 Q_j^2\right) \tilde{r}_{ijklm} u_{ij} Q_k Q_l Q_m - \tilde{z}_{ijklmn} u_{ij} Q_k Q_l Q_m Q_n + \widetilde{w}_{ijklmn} u_{ij} u_{kl} Q_m Q_n, \tag{S1.1d}$$

$$f_{grad} = \frac{1}{2} g_{ijkl} \frac{\partial Q_k}{\partial x_i} \frac{\partial Q_l}{\partial x_j}. \tag{S1.1e}$$

Here the subscripts $i, j, k, l, m, n$ … either designate 1) the phonon modes $\Gamma_{3-}$, $Y_{2+}$, and $Y_{4-}$, 2) the Cartesian indexes coupled to the strains $u_{ij}$, or 3) the Cartesian coordinates $x_i$. For instance, $Q_i$ stands for $Q_{\Gamma 3}$, $Q_{Y2}$, or $Q_{Y4}$. The summation rule is performed over repeated subscripts. Following Ref.[65], we pay special attention to the strong trilinear coupling of the $Q_{\Gamma 3}$, $Q_{Y2}$, and $Q_{Y4}$ amplitudes, where the energy $f_{tr}$ is proportional to the product $Q_{\Gamma 3} Q_{Y2} Q_{Y4}$, since the coupling can stabilize the FE o-phase.

Nonzero components of $\beta_i$, $\gamma$, $\delta_{ij}$, $\epsilon_i$, $\eta_{ijk}$, $\zeta_{ij}$, and $\xi_{ijkl}$ used in calculations for $HfO_2$ are listed in **Tables S1-S2** in **Appendix S1** [17]. $c_{ijkl}$ are elastic stiffnesses, $u_{ij}$ are elastic strains, $\tilde{q}_{ijkl}$, $\tilde{z}_{ijklmn}$, and $\widetilde{w}_{ijklmn}$ are the components of the second-order and higher-order striction *stress* tensors; $\tilde{r}_{ijklm}$ are the tensor of trilinear striction; and $g_{ijkl}$ are the components of the gradient energy tensor.

The ferroelectric polarization components $P_i$, both spontaneous and induced by external electric field $E_i^e$, contribute to the electric energy $f_{el}$:

$$f_{el} = -P_i\left(E_i^e + \frac{1}{2}E_i^d\right). \quad \text{(S1.2a)}$$

Here the subscript $i$ =1, 2, 3 and $E_i^d$ is the depolarization field, which should be determined from electrostatic equations in a self-consistent way. Following Refs. [15, 66, 67], the ferroelectric polarization $\vec{P}_3$ is proportional to the amplitude $Q_{\Gamma 3}$ of the $\Gamma_{3-}$ mode:

$$P_3 = \frac{Z_B^* d}{V_{f.u.}} Q_{\Gamma 3} \approx P_0 Q_{\Gamma 3}, \quad \text{(S1.2b)}$$

where $Z_B^*$ is the effective Bader charge [47], $V_{f.u.}$ is the formula unit (f.u.) volume, and $d$ is the elementary displacement corresponding to the polar $\Gamma_{3-}$ mode. The amplitude of the maximal atomic displacement $d$ was reported as 0.284, 0.278, and 0.268 Å for the polar, nonpolar, and antipolar modes, respectively [15].

Substitution of $P_3$ in Eqs.(S1.1) instead of $Q_{\Gamma 3}$, leads to the rescaling of the coefficients $\beta_i$, $\gamma$, $\delta_{ij}$, $\epsilon_i$, $\eta_{ijk}$, $\zeta_{ij}$, $\xi_{ijkl}$, $q_{ijkl}$, and $g_{ijkl}$ proportional to the powers of the factor $\frac{V_{f.u.}}{Z_B^* d} \approx \frac{1}{P_0}$. In particular, the biquadratic 2-4-6-8 coupling coefficients $\tilde{\beta}_{\Gamma 3}$, $\tilde{\delta}_{\Gamma 3}$, $\tilde{\eta}_{\Gamma 3}$, $\tilde{\xi}_{\Gamma 3}$ and the trilinear coupling coefficient $\tilde{\gamma}$, responsible for the behavior of the polar mode $\Gamma_{3-}$, have the form:

$$\tilde{\beta}_{\Gamma 3} = \frac{\beta_{\Gamma 3}}{P_0^2}, \quad \tilde{\delta}_{\Gamma 3} = \frac{\delta_{\Gamma 3}}{P_0^4}, \quad \tilde{\eta}_{\Gamma 3} = \frac{\eta_{\Gamma 3}}{P_0^6}, \quad \tilde{\xi}_{\Gamma 3} = \frac{\xi_{\Gamma 3}}{P_0^8}, \quad \tilde{\gamma} = \frac{\gamma}{P_0}. \quad \text{(S1.3)}$$

## APPENDIX S2. Free energy transformation from fixed strain to zero stress coefficients

Following the first-principles calculation of Delodovici et al. [15] we consider the phenomenological free energy at fixed values of strain in the following form:

$$\Delta F_{FE} = \zeta\big[\beta_\Gamma \Gamma^2 + \delta_\Gamma \Gamma^4 + \eta_\Gamma \Gamma^6 + \xi_\Gamma \Gamma^8 + \big(\gamma_{111} - (r_{13}u_1 + r_{23}u_2 + r_{33}u_3)\big)\Gamma\Psi\Phi + \beta_\Psi \Psi^2 + \delta_\Psi \Psi^4 + \eta_\Psi \Psi^6 + \xi_\Psi \Psi^8 + \beta_\Phi \Phi^2 + \delta_\Phi \Phi^4 + \eta_\Phi \Phi^6 + \xi_\Phi \Phi^8 + \delta_{\Gamma\Psi}\Psi^2\Gamma^2 + \delta_{\Psi\Phi}\Psi^2\Phi^2 + \delta_{\Gamma\Phi}\Phi^2\Gamma^2 + \epsilon_{\Gamma\Psi\Phi}\Gamma^3\Psi\Phi + \eta_{\Gamma\Psi\Psi}\Gamma^2\Psi^4 + \eta_{\Gamma\Gamma\Psi}\Gamma^4\Psi^2 + \eta_{\Gamma\Phi\Phi}\Gamma^2\Phi^4 + \eta_{\Gamma\Gamma\Phi}\Gamma^4\Phi^2 + \eta_{\Gamma\Psi\Phi}\Gamma^2\Psi^2\Phi^2 + \xi_{\Gamma\Psi}\Gamma^4\Psi^4 + \xi_{\Phi\Psi}\Phi^4\Psi^4 + \xi_{\Gamma\Phi}\Gamma^4\Phi^4\big] - \left(E_3^e + \frac{E_3^d}{2}\right)P_0\Gamma - (q_{13}u_1 + q_{23}u_2 + q_{33}u_3)P_0^2\Gamma^2 - (y_{133}u_1 + y_{233}u_2 + y_{333}u_3)P_0^4\Gamma^4 + \frac{1}{2}(c_{11}u_1^2 + c_{22}u_2^2 + c_{33}u_3^2) + (c_{12}u_1u_2 + c_{13}u_1u_3 + c_{23}u_2u_3) + \frac{1}{2}(c_{66}u_6^2 + c_{55}u_5^2 + c_{44}u_4^2). \quad \text{(S2.1)}$$

Here $\Gamma = Q_{\Gamma 3}$, $\Psi = Q_{Y2}$, and $\Phi = Q_{Y4}$ are the shortened re-designations for dimensionless polar, antipolar, and nonpolar order parameters, respectively (see **Appendix S1**). Note that the LGD expansion coefficients, which depend on the strain tensor, are used in Eq.(S2.1).

In Eq.(S2.1), we introduced the scale factor $\zeta = e/V_{f.u.}$ to transform from the atomic to SI units ($e$ is the elementary charge and $V_{f.u.}$ is the volume of the elementary cell). Voight notations are used in Eq.(S2.1):

$$c_{1111} = c_{11},\ \ c_{1122} = c_{12},\ \ c_{1212} = c_{44}, \tag{S2.2a}$$
$$q_{1111} = q_{11},\ \ q_{1122} = q_{12}, \tag{S2.2b}$$
$$u_{11} = u_1,\ \ u_{22} = u_2,\ \ 2u_{12} = u_6. \tag{S2.2c}$$

Modified Hooke's law can be obtained from the relation $\sigma_{ij} = \partial(\Delta F_{FE})/\partial\, u_{ij}$:

$$\sigma_1 = \mathrm{c}_{11}u_1 + \mathrm{c}_{12}u_2 + \mathrm{c}_{13}u_3 - q_{13}P_0^2\Gamma^2 - y_{133}P_0^4\Gamma^4 - \zeta r_{13}\Gamma\Psi\Phi, \tag{S2.3a}$$
$$\sigma_2 = c_{12}u_1 + c_{22}u_2 + c_{23}u_3 - q_{23}P_0^2\Gamma^2 - y_{233}P_0^4\Gamma^4 - \zeta r_{23}\Gamma\Psi\Phi, \tag{S2.3b}$$
$$\sigma_3 = c_{13}u_1 + c_{23}u_2 + c_{33}u_3 - q_{33}P_0^2\Gamma^2 - y_{333}P_0^4\Gamma^4 - \zeta r_{33}\Gamma\Psi\Phi, \tag{S2.3c}$$
$$c_{44}u_4 = \sigma_4, \quad c_{55}u_5 = \sigma_5, \quad c_{66}u_6 = \sigma_6. \tag{S2.3d}$$

The stress tensor components have the following form for the mechanically free system under hydrostatic pressure:

$$\sigma_1 = \sigma_2 = \sigma_3 = -p, \quad \sigma_4 = \sigma_5 = \sigma_6 = 0. \tag{S2.4}$$

Considering Eqs.(S2.3) and (S2.4), one could find the strain components as follows:

$$u_1 = -p(s_{11} + s_{12} + s_{13}) + Q_{13}P_0^2\Gamma^2 + Y_{133}P_0^4\Gamma^4 + \zeta Z_{13}\Gamma\Psi\Phi, \tag{S2.5a}$$
$$u_2 = -p(s_{22} + s_{12} + s_{23}) + Q_{23}P_0^2\Gamma^2 + Y_{233}P_0^4\Gamma^4 + \zeta Z_{23}\Gamma\Psi\Phi, \tag{S2.5b}$$
$$u_3 = -p(s_{13} + s_{23} + s_{33}) + Q_{33}P_0^2\Gamma^2 + Y_{333}P_0^4\Gamma^4 + \zeta Z_{33}\Gamma\Psi\Phi, \tag{S2.5c}$$
$$u_4 = u_5 = u_6 = 0. \tag{S2.5d}$$

Here we introduced the second-order electrostriction strain coefficients,

$$Q_{13} = s_{11}q_{13} + s_{12}q_{23} + s_{13}q_{33}, \tag{S2.6a}$$
$$Q_{23} = s_{12}q_{13} + s_{22}q_{23} + s_{23}q_{33}, \tag{S2.6b}$$
$$Q_{33} = s_{13}q_{13} + s_{23}q_{23} + s_{33}q_{33}, \tag{S2.6c}$$

the fourth-order electrostriction coefficients,

$$Y_{133} = s_{11}y_{133} + s_{12}y_{233} + s_{13}y_{333}, \tag{S2.7d}$$
$$Y_{233} = s_{12}y_{133} + s_{22}y_{233} + s_{23}y_{333}, \tag{S2.7d}$$
$$Y_{333} = s_{13}y_{133} + s_{23}y_{233} + s_{33}y_{333}, \tag{S2.7d}$$

and the trilinear striction

$$Z_{13} = s_{11}r_{13} + s_{12}r_{23} + s_{13}r_{33}, \tag{S2.8d}$$
$$Z_{23} = s_{12}r_{13} + s_{22}r_{23} + s_{23}r_{33}, \tag{S2.8d}$$
$$Z_{33} = s_{13}r_{13} + s_{23}r_{23} + s_{33}r_{33}. \tag{S2.8d}$$

It is easy to show that $r_{13}Q_{13} + r_{23}Q_{23} + r_{33}Q_{33} \equiv q_{13}Z_{13} + q_{23}Z_{23} + q_{33}Z_{33}$.

In Eqs.(S2.6)-(S2.8), we introduced the compliance tensor $s_{ij}$ as

$$s_{11} = \frac{c_{23}^2 - c_{22}c_{33}}{c_{13}^2c_{22} - 2c_{12}c_{13}c_{23} + c_{12}^2c_{33} + c_{11}(c_{23}^2 - c_{22}c_{33})},\ s_{12} = \frac{-c_{13}c_{23} + c_{12}c_{33}}{c_{13}^2c_{22} - 2c_{12}c_{13}c_{23} + c_{12}^2c_{33} + c_{11}(c_{23}^2 - c_{22}c_{33})}, \tag{S2.9a}$$

$$s_{13} = \frac{c_{13}c_{22} - c_{12}c_{23}}{c_{13}^2c_{22} - 2c_{12}c_{13}c_{23} + c_{12}^2c_{33} + c_{11}(c_{23}^2 - c_{22}c_{33})},\ s_{22} = \frac{c_{13}^2 - c_{11}c_{33}}{c_{13}^2c_{22} - 2c_{12}c_{13}c_{23} + c_{12}^2c_{33} + c_{11}(c_{23}^2 - c_{22}c_{33})}, \tag{S2.9a}$$

$$s_{23} = \frac{-c_{12}c_{13}+c_{11}c_{23}}{c_{13}^2c_{22}-2c_{12}c_{13}c_{23}+c_{12}^2c_{33}+c_{11}\left(c_{23}^2-c_{22}c_{33}\right)}, s_{33} = \frac{c_{12}^2-c_{11}c_{22}}{c_{13}^2c_{22}-2c_{12}c_{13}c_{23}+c_{12}^2c_{33}+c_{11}\left(c_{23}^2-c_{22}c_{33}\right)}. \quad \text{(S2.9a)}$$

Neglecting higher-order striction terms, the equation of state for polarization could be found from the minimization of Eq.(S2.1) with respect to Γ:

$$\zeta(2\beta_\Gamma\Gamma + 4\delta_\Gamma\Gamma^3 + 6\eta_\Gamma\Gamma^5 + 8\xi_\Gamma\Gamma^7 + [\gamma_{111} - (r_{13}u_1 + r_{23}u_2 + r_{33}u_3)]\Psi\Phi + 2[\delta_{\Gamma\Psi}\Psi^2 + \delta_{\Gamma\Phi}\Phi^2 + \eta_{\Gamma\Psi\Psi}\Psi^4 + \eta_{\Gamma\Phi\Phi}\Phi^4 + \eta_{\Gamma\Psi\Phi}\Psi^2\Phi^2]\Gamma + 3\zeta\epsilon_{\Gamma\Psi\Phi}\Gamma^2\Psi\Phi + 4[\eta_{\Gamma\Gamma\Psi}\Psi^2 + \eta_{\Gamma\Gamma\Phi}\Phi^2 + \xi_{\Gamma\Psi}\Psi^4 + \xi_{\Gamma\Phi}\Phi^4]\Gamma^3 + \cdots\Gamma^5) - 2(q_{13}u_1 + q_{23}u_2 + q_{33}u_3)P_0^2\Gamma = E_3^d + E_3^e. \quad \text{(S2.10a)}$$

After the substitution of the strain components from Eq.(S2.5) to Eq.(S2.10a) one obtains the expressions for the renormalized equation of state:

$$2[\zeta\beta_\Gamma + (Q_{12} + Q_{23} + Q_{33})P_0^2p]\Gamma + [4\zeta\delta_\Gamma - 2(q_{13}Q_{13} + q_{23}Q_{23} + q_{33}Q_{33})P_0^4]\Gamma^3 + 6\zeta\eta_\Gamma\Gamma^5 + 8\zeta\xi_\Gamma\Gamma^7 + \zeta[\gamma + (Z_{12} + Z_{23} + Z_{33})p]\Psi\Phi + \cdots + 3\zeta(\epsilon_{\Gamma\Psi\Phi} - [r_{13}Q_{13} + r_{23}Q_{23} + r_{33}Q_{33}]P_0^2)\Gamma^2\Psi\Phi + (2\zeta\eta_{\Gamma\Psi\Phi} - \zeta^2[r_{13}Z_{13} + r_{23}Z_{23} + r_{33}Z])\Gamma\Psi^2\Phi^2 = E_3^d + E_3^e. \quad \text{(S2.10b)}$$

## APPENDIX S3. The internal stress induced by the chemical strains in the shell

Strong chemical strains in the shell generate an elastic mismatch between the shell and the core, which induces stresses in the core. Below we calculate the stress induced by the linearized chemical strains (e.g., Vegard strains) in the paraelectric core.

It is convenient to rewrite the modified Hooke's law Eq.(S2.3) as follows:

$$u_1 = s_{11}\sigma_1 + s_{12}\sigma_2 + s_{13}\sigma_3 + Q_{13}P_0^2\Gamma^2 + \zeta Z_{13}\Gamma\Psi\Phi, \quad \text{(S3.1a)}$$
$$u_2 = s_{12}\sigma_1 + s_{22}\sigma_2 + s_{23}\sigma_3 + Q_{23}P_0^2\Gamma^2 + \zeta Z_{23}\Gamma\Psi\Phi, \quad \text{(S3.1b)}$$
$$u_3 = s_{13}\sigma_1 + s_{23}\sigma_2 + s_{33}\sigma_3 + Q_{33}P_0^2\Gamma^2 + \zeta Z_{33}\Gamma\Psi\Phi, \quad \text{(S3.1c)}$$
$$u_4 = s_{44}\sigma_4, \quad u_5 = s_{55}\sigma_5, \quad u_6 = s_{66}\sigma_6. \quad \text{(S3.1d)}$$

To find the elastic fields analytically, we use a perturbation approach. As a first step, we consider an isotropic elastic problem with spherical symmetry, consistent with the cubic symmetry of the paraelectric core-shell nanoparticle embedded in a soft-matter matrix. The elastic displacement in a spherical coordinate frame is given by expression, $\boldsymbol{U} = \{U_r(r), U_\theta(r), U_\phi(r)\}$, where $U_\theta = U_\phi = 0$ for a spherically symmetric case. In this case, the displacement vector satisfies the equation [68]:

$$\text{grad}(div\boldsymbol{U}) \equiv \frac{\partial}{\partial r}\frac{1}{r^2}\frac{\partial}{\partial r}(r^2U_r) = 0. \quad \text{(S3.2)}$$

It is seen from Eq.(S3.2), that the following relation takes place, $\left(\frac{\partial U_r}{\partial r} + 2\frac{U_r}{r}\right) = C_1$, and therefore the general solution of Eq.(S3.2) in the particle core ("c") and shell ("s") is:

$$U_r^c = C_1 r, \quad U_r^s = C_2 r + \frac{C_3}{r^2}. \quad \text{(S3.3)}$$

The strain tensor components are $u_{rr}^{c,s} = \frac{\partial U_r^{c,s}}{\partial r}$ and $u_{\theta\theta}^{c,s} = u_{\phi\phi}^{c,s} = \frac{U_r^{c,s}}{r}$, and their explicit form is

$$u_{rr}^c = u_{\theta\theta}^c = u_{\phi\phi}^c = C_1, \quad u_{rr}^s = C_2 - 2\frac{C_3}{r^3}, \quad u_{\theta\theta}^s = u_{\phi\phi}^s = C_2 + \frac{C_3}{r^3}. \quad \text{(S3.4)}$$

Substituting the solution (S3.4) into the Hooke's law (S3.1), we obtain the following expressions for the radial stresses:

$$\sigma_{rr}^{c}=\frac{C_1-q_cP_3^2-\zeta z_c\Gamma\Psi\Phi}{s_{11}^c+2s_{12}^c}, \tag{S3.5}$$

$$\sigma_{rr}^{s}=\frac{C_2-w_s-q_sP^2}{s_{11}^s+2s_{12}^s}-2\frac{C_3}{r^3}\frac{1}{s_{11}^s-s_{12}^s}\approx\frac{C_2-w_s}{s_{11}^s+2s_{12}^s}-2\frac{C_3}{r^3}\frac{1}{s_{11}^s-s_{12}^s}. \tag{S3.6}$$

Here $s_{ij}^c$ are elastic compliances, $q_c=(Q_{13}^c+Q_{23}^c+Q_{33}^c)/3$ and $z_c=(Z_{13}^c+Z_{23}^c+Z_{33}^c)/3$ are the isotropic parts of electrostriction tensors of the core; $s_{ij}^s$ are elastic compliances, $q_s=(Q_{13}^s+Q_{23}^s+Q_{33}^s)/3$ is an isotropic part of the electrostriction tensor, and $w_s$ are the Vegard strains of the shell. When deriving Eq.(S3.11), we neglected the higher order electrostriction contribution and assumed that the electric field and polarization in the core are homogeneous and directed along the polar axis "3". However, an inhomogeneous stray electric field can exist in the shell, and therefore we consider the total polarization of the shell to be $P^2=P_1^2+P_2^2+P_3^2$. Since the screening length $\lambda_{eff}$ of the shell is small (less than 1 nm) we can neglect the stray field, and thus omit the electrostriction term, $q_sP^2$, in the approximate equality in Eq. (S3.6).

The boundary conditions to Eq. (S3.2) are the continuity of radial elastic displacement and normal stress at core-shell interface, $r=R_c$,

$$u_r^c(R_c)=u_r^s(R_c),\quad \sigma_{rr}^c(R_c)=\sigma_{rr}^s(R_c), \tag{S3.7}$$

and the condition of a fixed pressure/tension at the shell surface, $r=R_s$,

$$\sigma_{rr}^s(R_s)=-p, \tag{S3.8}$$

where $p$ is an external pressure or intrinsic surface tension. The application of the boundary conditions (S3.7) and (S3.8) to the solution (S3.3)-(S3.6) yields the system of equations for the constants $C_i$:

$$C_1R_c=C_2R_c+\frac{C_3}{R_c^2}, \tag{S3.9a}$$

$$\frac{C_2-w_s}{s_{11}^s+2s_{12}^s}-2\frac{C_3}{R_c^3}\frac{1}{s_{11}^s-s_{12}^s}=\frac{C_1-q_cP_3^2-\zeta z_c\Gamma\Psi\Phi}{s_{11}^c+2s_{12}^c}, \tag{S3.9b}$$

$$\frac{C_2-w_s}{s_{11}^s+2s_{12}^s}-2\frac{C_3}{R_s^3}\frac{1}{s_{11}^s-s_{12}^s}=-p. \tag{S3.9c}$$

For the sake of simplicity below we assume that the elastic compliances of the core and the shell are the same: $s_{11}^s=s_{11}^c=s_{11}$ and $s_{12}^s=s_{12}^c=s_{12}$. In this case, the solution of Eqs.(S3.9) is:

$$C_1=q_cP_3^2+\zeta z_c\Gamma\Psi\Phi-\frac{s_{11}+2s_{12}}{s_{11}+s_{12}}\frac{2(R_s^3-R_c^3)}{3R_s^3}(q_cP_3^2+\zeta z_c\Gamma\Psi\Phi-w_s)-(s_{11}+2s_{12})p, \tag{S3.10a}$$

$$C_2=w_s+\frac{s_{11}+2s_{12}}{s_{11}+s_{12}}\frac{2R_c^3}{3R_s^3}(q_cP_3^2+\zeta z_c\Gamma\Psi\Phi-w_s)-(s_{11}+2s_{12})p, \tag{S3.10b}$$

$$C_3=\frac{s_{11}-s_{12}}{s_{11}+s_{12}}\frac{R_c^3}{3}(q_cP_3^2+\zeta z_c\Gamma\Psi\Phi-w_s). \tag{S3.10c}$$

The strain tensor components are:

$$u_{rr}^c = u_{\theta\theta}^c = u_{\phi\phi}^c = q_c P_3^2 + \zeta z_c \Gamma\Psi\Phi - \frac{s_{11}+2s_{12}}{s_{11}+s_{12}} \frac{2(R_s^3 - R_c^3)}{3R_s^3} (q_c P_3^2 + \zeta z_c \Gamma\Psi\Phi - w_s) - (s_{11} + 2s_{12})p, \tag{S3.11}$$

$$u_{rr}^s = w_s + \left(\frac{s_{11}+2s_{12}}{s_{11}+s_{12}} \frac{2R_c^3}{3R_s^3} - \frac{s_{11}-s_{12}}{s_{11}+s_{12}} \frac{2R_c^3}{3r^3}\right)(q_c P_3^2 + \zeta z_c \Gamma\Psi\Phi - w_s) - (s_{11} + 2s_{12})p, \tag{S3.12}$$

$$u_{\theta\theta}^s = u_{\phi\phi}^s = w_s + \left(\frac{s_{11}+2s_{12}}{s_{11}+s_{12}} \frac{2R_c^3}{3R_s^3} + \frac{s_{11}-s_{12}}{s_{11}+s_{12}} \frac{R_c^3}{3r^3}\right)(q_c P_3^2 + \zeta z_c \Gamma\Psi\Phi - w_s) - (s_{11} + 2s_{12})p. \tag{S3.13}$$

Non-trivial stress components in the core are

$$\sigma_{rr}^c = \sigma_{\theta\theta}^c = \sigma_{\phi\phi}^c = -p - \frac{2(R_s^3 - R_c^3)}{3R_s^3(s_{11}+s_{12})}(q_c P_3^2 + \zeta z_c \Gamma\Psi\Phi - w_s). \tag{S3.14}$$

It is seen that the stress tensor (S3.14) is equivalent to a hydrostatic pressure, normalized by the mismatch between the shell and the core.

For the case $p = 0$, Eq.(S3.14) is simplified as:

$$\sigma_{rr}^c = \sigma_{\theta\theta}^c = \sigma_{\phi\phi}^c = -\frac{2(R_s^3 - R_c^3)}{3R_s^3} \frac{qP_3^2 + \zeta z_c \Gamma\Psi\Phi - w_s}{s_{11}+s_{12}} \approx -2\frac{\Delta R}{R_s} \frac{q_c P_3^2 + \zeta z_c \Gamma\Psi\Phi - w_s}{s_{11}+s_{12}}, \tag{S3.15}$$

where $\Delta R$ is the shell thickness. The approximate equality is valid for thin shells, $(R_s - R_c) \ll R_c$. The nondiagonal stresses are absent,

$$\sigma_{r\theta}^c = \sigma_{r\phi}^c = \sigma_{\phi\theta}^c = 0. \tag{S3.16}$$

In the important case of different elastic compliances of the core and the shell, the rather cumbersome solution of Eqs.(S3.9) can be found. Of particular interest is the stress in the core, which is given by the following expression:

$$\sigma_{rr}^c = \sigma_{\theta\theta}^c = \sigma_{\phi\phi}^c = \frac{-2(R_s^3 - R_c^3)(q_c P_3^2 + \zeta z_c \Gamma\Psi\Phi - w_s) - 3R_s^3(s_{11}^s + s_{12}^s)p}{R_s^3(2s_{11}^c + 4s_{12}^c + s_{11}^s - s_{12}^s) - 2R_c^3(s_{11}^c + 2s_{12}^c - s_{11}^s - 2s_{12}^s)}. \tag{S3.17}$$

After substitution of the stress (S3.14) into Eq.(S2.10b) and elementary transformations, which is equivalent to the substitution,

$$p \to p - \frac{2(R_s^3 - R_c^3)w_s}{3R_s^3(s_{11}+s_{12})} + \frac{2(R_s^3 - R_c^3)}{3R_s^3(s_{11}+s_{12})}\left(q_c P_0^2 \Gamma^2 + \zeta z_c \Gamma\Psi\Phi\right), \tag{S3.18}$$

the following equation for the spontaneous polarization with renormalized coefficients is obtained:

$$\left[2\zeta\beta_\Gamma + \frac{1}{\varepsilon_b + 2\varepsilon_s + (R_c/\lambda_{eff})} \frac{P_0^2}{\varepsilon_0} + 6q_c P_0^2 \left\{p - \frac{2(R_s^3 - R_c^3)w_s}{3R_s^3(s_{11}+s_{12})}\right\}\right]\Gamma + \left[4\zeta\delta_\Gamma - 2(q_{13}Q_{13} + q_{23}Q_{23} + q_{33}Q_{33})P_0^4 + \frac{2(q_c P_0^2)^2 (R_s^3 - R_c^3)}{R_s^3(s_{11}+s_{12})}\right]\Gamma^3 + \zeta\left[\gamma + 3z_c\left\{p - \frac{2(R_s^3 - R_c^3)w_s}{3R_s^3(s_{11}+s_{12})}\right\}\right]\Psi\Phi + \dots + 3\zeta\left(\epsilon_{\Gamma\Psi\Phi} - [r_{13}Q_{13} + r_{23}Q_{23} + r_{33}Q_{33}]P_0^2 + z_c q_c P_0^2 \frac{2(R_s^3 - R_c^3)}{R_s^3(s_{11}+s_{12})}\right)\Gamma^2\Psi\Phi + \left(2\zeta\eta_{\Gamma\Psi\Phi} - \zeta^2[r_{13}Z_{13} + r_{23}Z_{23} + r_{33}Z_{33}] + (\zeta z_c)^2 \frac{2(R_s^3 - R_c^3)}{R_s^3(s_{11}+s_{12})}\right)\Gamma\Psi^2\Phi^2 = E_3^e. \tag{S3.19}$$

Only renormalized terms are shown. The shell influence on the coefficients (S3.19) is presented by the factor $\frac{2(R_s^3 - R_c^3)}{3R_s^3}$. The parameter $\Delta_\beta$, introduced in Eq.(6) in the main text, can be estimated from Eq.(S3.19) as

$$\Delta_\beta \cong -\frac{\zeta\beta_\Gamma}{P_0^2} - \frac{f_m}{P_0^2}. \quad \text{(S3.20)}$$

## APPENDIX S4. Piezoelectric effect and electrostriction

Piezoelectric effect tensors were calculated for $HfO_2$ by Dutta et al. [69]. Electrostriction coefficients were calculated in Ref. [17]. The relationship between the piezoelectric ($d_{ijk}$) and electrostriction ($Q_{jkmn}$) tensors is

$$d_{ijk} = 2\varepsilon_0\varepsilon_{im}P_nQ_{jkmn}, \quad \text{(S4.1)}$$

where $\varepsilon_0$ is the universal dielectric constant, $\varepsilon_{im}$ is the tensor of dielectric permittivity, and $P_n$ is the spontaneous polarization component. The explicit form of Eq.(S4.1) for the case of the o-symmetry is given in **Table S1.**

**Table S1**. Relations between the piezoelectric ($d_{ijk}$) and electrostriction ($Q_{jkmn}$) tensors

| Piezoelectric tensor components in matrix and Voight notations | Expression via the electrostriction coefficients |
|---|---|
| $d_{311} = d_{31}$ | $2\varepsilon_0\varepsilon_{33}P_3Q_{1133}$ |
| $d_{322} = d_{32}$ | $2\varepsilon_0\varepsilon_{33}P_3Q_{2233}$ |
| $d_{333} = d_{33},$ | $2\varepsilon_0\varepsilon_{33}P_3Q_{3333}$ |
| $d_{113} = d_{15}/2$ | $2\varepsilon_0\varepsilon_{11}P_3Q_{1313}$ |
| $d_{223} = d_{24}/2$ | $2\varepsilon_0\varepsilon_{22}P_3Q_{2323}$ |

Using the electrostriction coefficients calculated in Ref. [17], namely $Q_{1133} = -0.0139, Q_{2233} = -0.00853, Q_{3333} = -0.008423, Q_{1313} = -0.0241, Q_{2323} = 0.0847$ (in $m^4/C^2$ units); dielectric permittivity calculated by Zhao and Vanderbilt [70], namely $\varepsilon_{11}$ =23 and $\varepsilon_{22}$ =18; and the values of the average polarization $\bar{P}_s$ and permittivity $\varepsilon_{33} = \overline{\frac{dP_3}{dE_3^e}}$ calculated in this work, we recalculated the color maps of the piezoelectric coefficient $d_{333}$, in **Fig. 3(d)-(f)**, and the volumetric sum $d_{333} + d_{322} + d_{311}$, shown in **Fig. 3(g)-(i).**

## APPENDIX S5. Internal stresses induced in core-shell nanoparticles by inhomogeneous chemical strains in the shell

Chemical strains (e.g., Vegard strains) in the shell of the nanoparticle induce an elastic mismatch at the shell-core interface, which generates stresses in the nanoparticle. The stresses induced by chemical strains satisfy the modified Hooke's law Eq.(S2.3) in the paraelectric phase:

$$u_1 = s_{11}\sigma_1 + s_{12}\sigma_2 + s_{13}\sigma_3 + u_c, \quad \text{(S5.1a)}$$

$$u_2 = s_{12}\sigma_1 + s_{22}\sigma_2 + s_{23}\sigma_3 + u_c, \quad \text{(S5.1b)}$$

$$u_3 = s_{13}\sigma_1 + s_{23}\sigma_2 + s_{33}\sigma_3 + u_c, \tag{S5.1c}$$

$$u_4 = s_{44}\sigma_4, \quad u_5 = s_{55}\sigma_5, \quad u_6 = s_{66}\sigma_6. \tag{S5.1d}$$

Here, $s_{ij}$ are elastic compliances and $u_c$ is the isotropic chemical strain inside the particle. The components $U_i$ of displacement vector $\vec{U}$ satisfy Lame-type equations [71]:

$$\frac{s_{11}}{2(s_{11}-s_{12})(s_{11}+2s_{12})}\frac{\partial}{\partial x_i}\frac{\partial U_j}{\partial x_j} + \frac{1}{2(s_{11}-s_{12})}\frac{\partial}{\partial x_j}\frac{\partial U_i}{\partial x_j} - \frac{1}{s_{11}+2s_{12}}\frac{\partial u_c}{\partial x_i} = 0\ , \quad i\text{=1, 2, 3.} \tag{S5.2a}$$

For the sake of simplicity, we assumed that the tensor of elastic compliances of the particle is isotropic and thus the following relations are valid: $s_{11} = s_{22} = s_{33}$ and $s_{12} = s_{13} = s_{23}$. Using identical equalities

$$\frac{\partial}{\partial x_i}\frac{\partial U_j}{\partial x_j} \equiv \frac{\partial}{\partial x_j}\frac{\partial U_i}{\partial x_j} + e_{ilm}\frac{\partial}{\partial x_m}e_{ljk}\frac{\partial U_j}{\partial x_k}\ , \tag{S5.2b}$$

one can transform Eq.(S5.2a) in the following way

$$\frac{s_{11}+s_{12}}{s_{11}-s_{12}}\frac{\partial}{\partial x_i}\frac{\partial U_j}{\partial x_j} + \frac{s_{11}+2s_{12}}{2(s_{11}-s_{12})}\left(-e_{ilm}\frac{\partial}{\partial x_m}e_{ljk}\frac{\partial U_j}{\partial x_k}\right) - \frac{\partial u_c}{\partial x_i} = 0 \quad \Leftrightarrow$$

$$\frac{s_{11}+s_{12}}{s_{11}-s_{12}}\mathrm{grad}\left(\mathrm{div}\vec{U}\right) - \frac{s_{11}+2s_{12}}{2(s_{11}-s_{12})}\mathrm{rot}\left(\mathrm{rot}\vec{U}\right) - \mathrm{grad}(u_c) = 0. \tag{S5.2c}$$

To derive analytical expressions for the elastic fields, we consider an isotropic system with spherical symmetry, where all the variables depend only on the radial coordinate $r$. The elastic displacement has the following form in spherical coordinates: $\vec{U} = \{U_r(r), U_\theta(r), U_\phi(r)\}$, where $U_\theta = U_\phi = 0$ for a spherically symmetric case. Therefore, the Lame equation (S.2a) can be simplified as

$$\frac{\partial}{\partial r}\left[\frac{1}{r^2}\frac{\partial}{\partial r}(r^2 U_r)\right] = \frac{s_{11}-s_{12}}{s_{11}+s_{12}}\frac{\partial u_c}{\partial r}. \tag{S5.2d}$$

It is seen from Eq.(S5.2d), that the following relation takes place, $\frac{1}{r^2}\frac{\partial}{\partial r}(r^2 U_r) = \frac{s_{11}-s_{12}}{s_{11}+s_{12}}u_c + C_1$, and therefore the general solution of Eq.(S5.2d) is:

$$U_r = C_1 r + \frac{1}{r^2}\left[C_0 + \frac{s_{11}-s_{12}}{s_{11}+s_{12}}\int_0^r u_c(\rho)\rho^2 d\rho\right]. \tag{S5.3}$$

The strain tensor components in spherical coordinates are $u_{rr} = \frac{\partial U_r}{\partial r}$ and $u_{\theta\theta} = u_{\phi\phi} = \frac{U_r}{r}$. Their explicit form is

$$u_{rr} = C_1 - \frac{2}{r^3}\left[C_0 + \frac{s_{11}-s_{12}}{s_{11}+s_{12}}\int_0^r u_c(\rho)\rho^2 d\rho\right] + \frac{s_{11}-s_{12}}{s_{11}+s_{12}}u_c(r), \tag{S5.4a}$$

$$u_{\theta\theta} = u_{\phi\phi} = C_1 + \frac{1}{r^3}\left[C_0 + \frac{s_{11}-s_{12}}{s_{11}+s_{12}}\int_0^r u_c(\rho)\rho^2 d\rho\right]. \tag{S5.4b}$$

Substituting expressions (S5.4) into Hooke's law (S5.1), we obtain the following expressions for the stress components:

$$\sigma_{rr} = \frac{C_1}{s_{11}+2s_{12}} - \frac{2}{r^3}\frac{1}{s_{11}-s_{12}}\left[C_0 + \frac{s_{11}-s_{12}}{s_{11}+s_{12}}\int_0^r u_c(\rho)\rho^2 d\rho\right], \tag{S5.5a}$$

$$\sigma_{\theta\theta} = \sigma_{\phi\phi} = \frac{C_1}{s_{11}+2s_{12}} + \frac{1}{r^3}\frac{1}{s_{11}-s_{12}}\left[C_0 + \frac{s_{11}-s_{12}}{s_{11}+s_{12}}\int_0^r u_c(\rho)\rho^2 d\rho\right] - \frac{u_c(r)}{s_{11}+s_{12}}. \quad \text{(S5.5b)}$$

The integral $\int_0^r u_c(\rho)\rho^2 d\rho / r^3 \underset{r\to 0}{\longrightarrow} u_c(0)/3$ for the finite $u_c(0)$. The condition of finite $\sigma_{rr}$ at $r \to 0$ requires $C_0 = 0$. The condition of zero radial stress at the particle surface at $r = R$,

$$\sigma_{rr}(R) = 0, \quad \text{(S5.6)}$$

gives the relation $C_1 = \frac{2}{R^3}\frac{s_{11}+2s_{12}}{s_{11}+s_{12}}\int_0^R u_c(\rho)\rho^2 d\rho$. Finally, we could obtain the distribution of stresses:

$$\sigma_{rr} = \frac{2}{3(s_{11}+s_{12})}\left[\langle u_c\rangle - \frac{3}{r^3}\int_0^r u_c(\rho)\rho^2 d\rho\right], \quad \text{(S5.7a)}$$

$$\sigma_{\theta\theta} = \sigma_{\phi\phi} = \frac{1}{s_{11}+s_{12}}\left[\frac{2}{3}\langle u_c\rangle + \frac{1}{r^3}\int_0^r u_c(\rho)\rho^2 d\rho - u_c(r)\right], \quad \text{(S5.7b)}$$

$$\sigma_{rr} + \sigma_{\theta\theta} + \sigma_{\phi\phi} = \frac{2}{s_{11}+s_{12}}[\langle u_c\rangle - u_c(r)]. \quad \text{(S5.7c)}$$

The chemical strain $\langle u_c\rangle$, averaged over the nanoparticle volume and proportional to the total amount of the defects, is introduced in Eq.(S5.7):

$$\langle u_c\rangle = \frac{3}{R^3}\int_0^R u_c(\rho)\rho^2 d\rho. \quad \text{(S5.7d)}$$

Hereinafter we redesignate $R_s \to R$ and omit the subscript "s" for the sake of simplicity. According to Eqs.(S5.7), the stress in the center of the nanoparticle (at $r \ll R$) is given by expression:

$$\sigma_{rr} = \sigma_{\theta\theta} = \sigma_{\phi\phi} = \frac{2}{3(s_{11}+s_{12})}[\langle u_c\rangle - u_c(r)]. \quad \text{(S5.8)}$$

It is seen that the stress in Eq. (S5.8) is equivalent to the hydrostatic pressure induced by the difference of average and local chemical strains. Relative volume changes can be estimated as:

$$u_1 + u_2 + u_3 = (s_{11} + 2s_{12})(\sigma_1 + \sigma_2 + \sigma_3) + 3u_c = \frac{2(s_{11}+2s_{12})}{s_{11}+s_{12}}\langle u_c\rangle + \frac{s_{11}-s_{12}}{s_{11}+s_{12}}u_c(r). \quad \text{(S5.9)}$$

Below we consider several characteristic distributions of the defect-induced chemical strains $u_c(r)$ in Eqs.(S5.7) and calculate corresponding elastic stresses in the nanoparticle.

**Case I.** The distributions of defects and corresponding chemical strain are homogeneous, $u_c(r) = \langle u_c\rangle$, and so the internal stress is zero

$$\sigma_{rr} = \sigma_{\theta\theta} = \sigma_{\phi\phi} = 0. \quad \text{(S5.10)}$$

**Case II.** The distributions of defects and corresponding chemical strains are step-like and localized in the shell with thickness $\Delta R$ (see curve 1 in **Fig. S1(a)**):

$$u_c(r) = \begin{cases} 0, & r < R - \Delta R, \\ u_0, & R - \Delta R \le r \le R. \end{cases} \quad \text{(S5.11a)}$$

The integral $\frac{3}{r^3}\int_0^r u_c(\rho)\rho^2 d\rho = 0$ for $r < R - \Delta R$ and $\frac{u_0}{r^3}[r^3 - (R - \Delta R)^3]$ for $R - \Delta R \le r \le R$. The internal stress is

$$\sigma_{rr} = \sigma_{\theta\theta} = \sigma_{\phi\phi} = \frac{2u_0}{s_{11}+s_{12}}\left[\frac{R^3-(R-\Delta R)^3}{3R^3}\right] \quad \text{at} \quad r < R - \Delta R. \quad \text{(S5.11b)}$$

The stresses and strains are shown by curves 1 in **Fig. S1(b)-(d)**.

**Case III.** The defects and corresponding chemical strains are exponentially distributed (see curve 2 in **Fig. S1(a)**):

$$u_c(r) = u_0 \exp\left(-\frac{R-r}{\mathrm{d}}\right), \qquad 0 \le r \le R. \tag{S5.12a}$$

The integral

$$\frac{1}{r^3}\int_0^r u_c(\rho)\rho^2 d\rho = u_0 \frac{\mathrm{d}}{r^3} e^{\frac{r-R}{\mathrm{d}}}\left(2d^2\left[1-e^{-\frac{r}{d}}\right]-2dr+r^2\right). \tag{S5.12a}$$

The explicit form of strain and stress distributions is rather cumbersome and could be found from Eqs.(S5.4) and (S5.7). They are shown by curves 2 in **Fig. S1(b)-(d)**.

**Case IV.** The distributions of defects and corresponding chemical strains are Fermi-like (see curve 3 in **Fig. S1(a)**):

$$u_c(r) = \frac{u_0}{1+\exp\left(\frac{R-\Delta R-r}{\delta R}\right)}, \qquad 0 \le r \le R. \tag{S5.13}$$

The explicit form of strain and stress distributions has the form of quadratures, which can be found from Eqs.(S5.4) and (S5.7). They are shown by curves 3 in **Fig. S1(b)-(d)**.

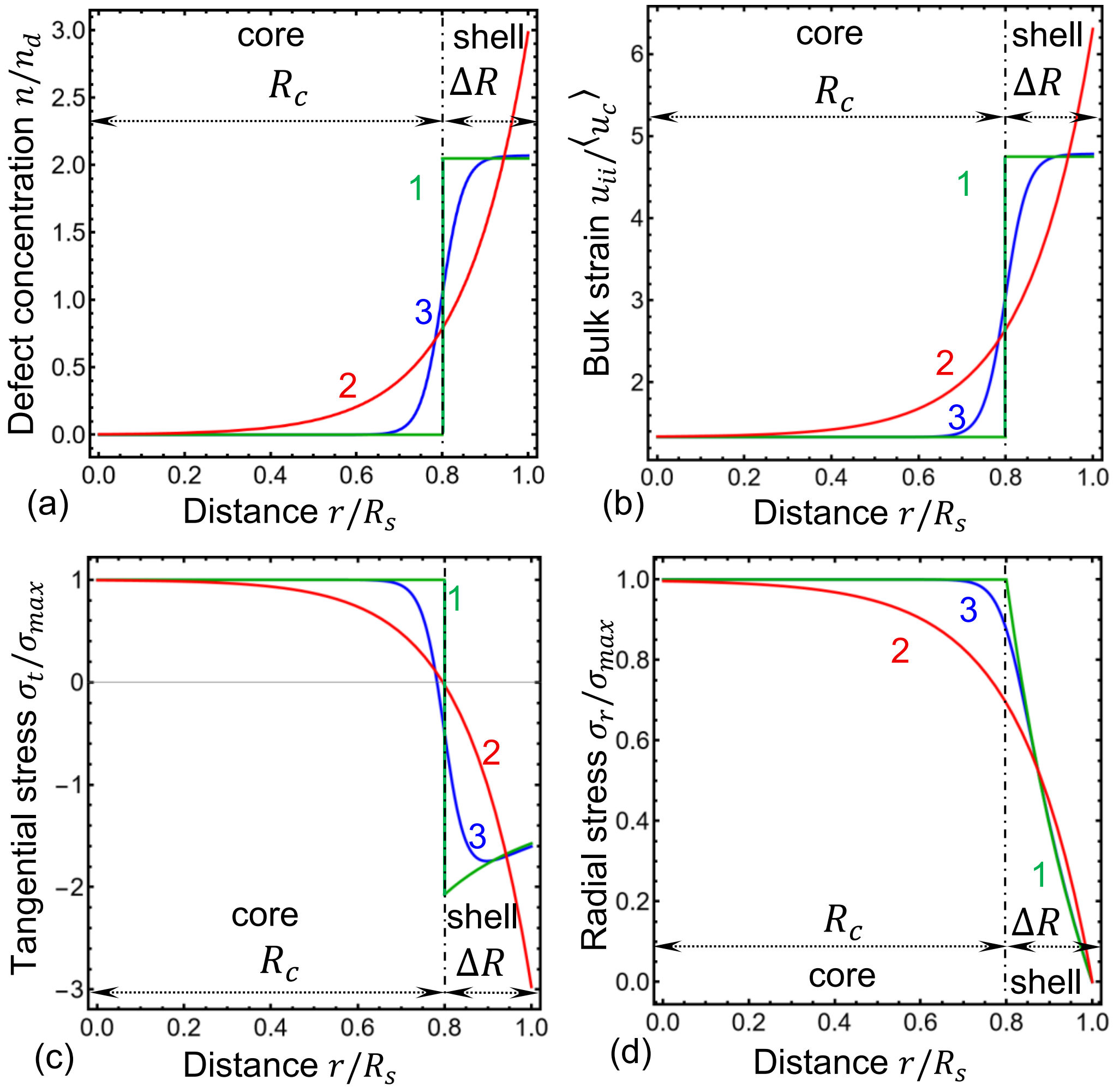

**Figure S1.** Radial distributions of the defect concentration, normalized to its average value $n_d$ **(a)**, the bulk strain, normalized to the average chemical strain $\langle u_c \rangle$ **(b)**, and the tangential **(c)** and the radial **(d)** stresses, normalized to their maximal values. Curves 1-3 in plots **(b)-(d)** are calculated for different defect concentration distribution functions: step-like (curves 1), exponential (curves 2), and Fermi-like (curves 3), corresponding to the distributionsshown in part **(a)**. The parameters are: $\Delta R/R_s = 0.2$, $d/R_s = 0.15$, and $\delta R/R_s = 0.025$. Note that the total number of defects, $N_d = 4\pi \int_0^R n(\rho)\rho^2 d\rho$, is conserved for all three distributions.

It can be seen from **Fig. S1(d)** that the radial stress is almost constant in the core for the Fermi-like defect distribution: the deviation of curve 3 from curve 1 does not exceed $5 - 10$ % at $r < R_c$. The deviation of the stress from its constant value does not exceed $10 - 25$ % for the exponential defect distribution (compare curves 1 and 2). This result justifies the step-like core-shell model for elastic defects used in this work.

Moreover, the FEM results confirm that the calculated values of the physical quantities, such as the core polarization and piezoelectric coefficients, differ very little (the deviation does not exceed 5 %) for the considered defect distributions, provided that the total number of defects, $N_d$, is conserved. The intuitive condition of constant $N_d$ has a relatively simple explanation based on the Saint-Venant principle and the direct variational method. Indeed, we consider the spatially averaged values of the free energy with renormalized coefficients instead of the local distributions of the elastic and electric fields (see, e.g., Ref. [72]). The integration of the elastic fields over the core volume minimizes the difference between the defect distribution functions, making the results primarily sensitive to the average defect concentration. Therefore, Eqs. (4)-(6), used in the main text, are robust with respect to the details of the defect distribution function and primarily sensitive to the average defect concentration $n_d$, shell thickness $\Delta R$ and core radius $R_c$, because their product is proportional to the total number of defects $N_d$. For thin shells ($R_c \gg \Delta R$), mainly considered in this work, the total number of defects, $N_d \approx \frac{4\pi}{3}(R_s^3 - R_c^3)n_d \approx 4\pi R_c^2 \Delta R n_d$, should be constant for the core-shell model to remain valid.